\documentclass{pasa}%

\usepackage{graphicx}
\usepackage{xcolor}
\usepackage{amsmath}	
\usepackage{amssymb}	
\usepackage{rotating}
\usepackage{supertabular,booktabs}
\usepackage{subcaption}

\usepackage{times}
\usepackage{color}
\usepackage{mathptmx}

\newcommand{\myaffil}[1]{$^{\rm #1}$}
\newcounter{inst}
\newcommand{\inst}[1]{\noindent%
   \refstepcounter{inst}\myaffil{\arabic{inst}\label{#1}}     
   }
   
\newcommand{\arcdeg}{\ensuremath{^{\circ}}}

\newcommand{\ncol}{388}
\newcommand{\survarea}{12,892}

\newcommand{\nsrc}{624,866}

\newcommand{\nfit}{562,302}

\newcommand{\ncplfit}{18,869}
\newcommand{\srcdensity}{48}
\newcommand{\pctreliablelow}{98.7}
\newcommand{\pctreliablehigh}{99.75}

\newcommand{\sect}{Section}

\newcommand{\Sect}{Section}

\newcommand{\Tab}{Table}

\newcommand{\dri}{GLEAM-X DRI}
\newcommand{\drii}{GLEAM-X DRII}

\newcommand{\dg}{\mbox{\ensuremath{^\circ}}}
\newcommand{\arcsec}{\mbox{\ensuremath{''}}} 

\title[GLEAM-X Second Data Release]{GaLactic and Extragalactic All-sky Murchison Widefield Array eXtended (GLEAM-X) survey II: Second Data Release}
\author[K.~Ross~et~al.]{K.~Ross\myaffil{\ref{ICRAR}},
N.~Hurley-Walker\myaffil{\ref{ICRAR}},
T.~J.~Galvin\myaffil{\ref{ICRAR},\ref{CASS}},
B.~Venville\myaffil{\ref{ICRAR}},
S.~W.~Duchesne\myaffil{\ref{CASS}},
J.~Morgan\myaffil{\ref{ICRAR},\ref{CASS}},
T.~An\myaffil{\ref{SHAO}},
G.~G\"{u}rkan\myaffil{\ref{Hertfordshire},\ref{CASS}}
P.~J.~Hancock\myaffil{\ref{ICRAR},\ref{CIDS}},
G.~Heald\myaffil{\ref{CASS}},
M.~Johnston-Hollitt\myaffil{\ref{CIDS}},
S.~V.~White\myaffil{\ref{ICRAR},\ref{SAAO}}\\
{\small \myaffil{}\,Email: kathryn.ross@icrar.org}\\
{\small \inst{ICRAR}\,International Centre for Radio Astronomy Research, Curtin University, Bentley, WA 6102, Australia}\\
{\small \inst{CASS}\,CSIRO Space \& Astronomy, PO Box 1130, Bentley WA 6102, Australia}\\
{\small \inst{CIDS}\,Curtin Institute for Data Science, Curtin University, GPO Box U1987, Perth WA 6845, Australia}\\
{\small \inst{SHAO}\,Shanghai Astronomical Observatory, Chinese Academy of Sciences, 80 Nandan Rd, Shanghai, 200030, China}\\
{\small \inst{Hertfordshire}\,Centre for Astrophysics Research, University of Hertfordshire, College Lane, Hatfield AL10 9AB, UK}\\
{\small \inst{SAAO}\,South African Astronomical Observatory, PO Box 9, Observatory, 7935, South Africa}\\
}

\jid{PASA}
\doi{10.1017/pas.\the\year.xxx}
\jyear{\the\year}

\usepackage{aas_macros}
\usepackage{hyperref} 
\hypersetup{colorlinks,citecolor=blue,linkcolor=blue,urlcolor=blue}

\begin{document}

\begin{frontmatter}
\maketitle

\begin{abstract}
We present the second data release for the GaLactic and Extragalactic All-sky Murchison Widefield Array eXtended (GLEAM-X) survey. This data release is an area of \survarea{-deg$^2$} around the South Galactic Pole region covering 20\,h40\,m$\leq$RA$\leq$6\,h40\,m, -90\dg{}$\leq$Dec$\leq$+30\dg{}. Observations were taken in 2020 using the Phase-II configuration of the Murchison Widefield Array (MWA) and covering a frequency range of 72--231\,MHz with twenty frequency bands. We produce a wideband source-finding mosaic over 170--231\,MHz with a median root mean squared noise of $1.5^{+1.5}_{-0.5}$\,mJy beam$^{-1}$. We present a catalogue of \nsrc{} components, including \nfit{} components which are spectrally fit. This catalogue is 98\% complete at 50\,mJy, and a reliability of 98.7\% at a 5\,$\sigma$ level, consistent with expectations for this survey. The catalogue is made available via Vizier and the PASA datastore and accompanying mosaics for this data release are made available via AAO Data Central and SkyView. 
\end{abstract}

\begin{keywords}
techniques: interferometric -- galaxies: general -- radio continuum: surveys
\end{keywords}

\end{frontmatter}



\section{Introduction}
Wide-area radio sky surveys enable a range of science across the Universe, from the most nearby scales, e.g. measuring the solar wind \citep{2023SpWea..2103396M}, to our own Galaxy, e.g. finding unexpected types of transient radio sources \citep{2022Natur.601..526H}, to the extragalactic sky \citep{2020PASA...37...17W,2020PASA...37...18W}, such as measurements and variability of radio galaxies \citep{2022MNRAS.512.5358R} and galaxy clusters \citep{2020PASA...37...37D}, the detection of cosmic magnetism \citep{2021MNRAS.505.4178V}, precision magnetism studies \citep{2018PASA...35...43R,2020PASA...37...29R}, and ultimately cosmology \citep{2024MNRAS.527.6540H}. 

Building toward the Square Kilometre Array (SKA), a recent resurgence in low-frequency radio astronomy has seen a plethora of new wide-area radio sky surveys. \citet{2022PASA...39...35H} (hereafter referred to as HW22) introduced the GaLactic and Extragalactic All-Sky MWA -- eXtended (GLEAM-X) survey, using the Murchison Widefield Array \citep[MWA;][]{2013PASA...30....7T} in its ``extended'' Phase-\textsc{ii} configuration \citep{2018PASA...35...33W,2019PASA...36...50B} to survey the sky south of Declination~$+30^\circ$ over 72--231\,MHz, with a source-finding image formed at $\sim56''$ resolution. The authors also introduced the many other wide-area surveys that formed the foundation for, were contemporaneous with, or preceded GLEAM-X. Adding to the recent plethora of data releases, this paper presents the second data release to the GLEAM-X survey, henceforth called GLEAM-X~DRII, which will add to the scientific resources available to the astronomical community. 

Since GLEAM-X was described, a multitude of new radio surveys of the sky have become available or have had new data releases. The LOw-Frequency Array \citep[LOFAR;][]{2013A&A...556A...2V} has produced a second data release from the LOFAR Two-metre Sky Survey at 120--180\,MHz \citep[LoTSS;][]{2017A&A...598A.104S, 2022A&A...659A...1S}, a first data release from the LOFAR Low-Band Array Sky Survey at 41--66\,MHz \citep[LoLSS;][]{2021A&A...648A.104D, 2023A&A...673A.165D}, and deep imaging toward the LOFAR Deep Fields \citep{2023MNRAS.523.1729B}. Dish interferometers with phased-array feeds have enabled wide-area higher-frequency surveys, such as the Apertif imaging survey \citep{2022A&A...667A..38A}, and the Rapid Australian SKA Pathfinder Continuum Surveys (RACS), in both unpolarised and circular polarisation \citep{2020PASA...37...48M,2023PASA...40...34D}, as well as early results from measuring linear polarisation \citep{2023PASA...40...40T}. MeerKAT has also produced L-band surveys over smaller areas at greater sensitivities, such as the MeerKAT Absorption Line Survey (MALS) data release~\textsc{i} \citep[MALS;][]{2023arXiv230812347D} and MIGHTEE \citep{2022MNRAS.509.2150H}. The first data release from the MWA interlpanetary scintillation (IPS) survey has also been released \citep{2022PASA...39...63M} which provides information on the sub-arcsecond structure of over 40,000 GLEAM sources. 

While GLEAM-X is one of the several ongoing low-frequency surveys, it continues to play a unique role combining wide fractional bandwidth, low frequencies, and large sky coverage, particularly in the Southern sky. Covering Declinations up to $+30^\circ$, GLEAM-X is complimentary to Northern sky surveys, like LoTSS, and bridges the gap as a Southern sky survey until SKA-LOW. Likewise, the synergies between other higher-frequency Southern sky surveys like RACS will prove a powerful source for discovery and characterisation of spectral properties for large populations of radio sources. The increased sensitivity and resolution of GLEAM-X compared to the predecessor, GLEAM, establish GLEAM-X as an avenue for novel scientific outcomes that were not available with GLEAM alone. 

This paper is presented with the following layout. In Section~\ref{sec:obs}, we describe the observational strategy and specific observations processed for this data release. In Section~\ref{sec:dr}, we outline the changes and improvements made to the GLEAM-X processing pipeline since the initial data release and survey description \citep{2022PASA...39...35H}. The properties of the final images are outlined in Section~\ref{sec:images}. In Section~\ref{sec:catalogue}, we describe the compact source catalogue produced for this work and quality of this catalogue. All the results are summarised in Section~\ref{sec:summary}.
 
All positions given in this paper are in J2000 equatorial coordinates.

\section{Observations}\label{sec:obs}

GLEAM-X utilised the drift-scan observing technique, outlined by \citet{2015PASA...32...25W}, but adapted to iterate over three hour angles (HA), $HA=0$\,h,$\pm1$\,h, proven to increase sensitivity \citep{2021PASA...38...14F}. \dri{} comprised four drift scans at Declinations centred on $-26^\circ$ and covering $4\mathrm{h}\leq \mathrm{RA}\leq13$\,h with different HAs \citep{2022PASA...39...35H}. Observations for \drii{} adopt the same observing strategy and cover the South Galactic Pole (SGP) region, spanning $\sim 20$\,h$\leq\mathrm{RA}\leq \sim 6$\,h and Dec$\leq+30$. \drii{} consists of 28 drift scans, with seven Declination pointings ($+20^\circ$, $+1^\circ$, $-12^\circ$, $-26^\circ$, $-40^\circ$, $-55^\circ$, and $-71^\circ$) and $\mathrm{HAs}=0$\,h,$\pm1$\,h. Each Declination pointing had four dedicated observing nights separated by one week and were taken from 2020-09-28 to 2020-10-25. A summary of the nights that were included in this data release is presented in \Tab~\ref{tab:obs}.

\begin{figure}
    \centering
    \includegraphics[width=1\linewidth]{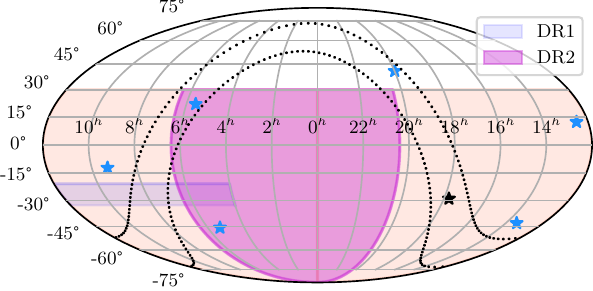}
    \caption{Sky coverage of the GLEAM-X survey. The blue region represents the area covered in the first data release (HW22), the region covered as part of this release is covered by the pink region and the total coverage of the GLEAM-X survey is shown by the cream region. The black star represents the Galactic centre and the black dotted lines represent the Galactic plane from -10\dg$\leq b \leq$10\dg. The blue stars represent the bright A-team sources: Centaurus A, Crab, Cygnus A, Hydra A, Pictor A, and Virgo A.}
    \label{fig:gleamx_sky_coverage}
\end{figure}

As outlined by HW22, observations were taken with an instantaneous bandwidth of $30.72$-MHz and cycled through five frequency ranges, 72--103\,MHz, 103--134\,MHz, 139--170\,MHz, 170--200\,MHz, and 200--231\,MHz, every two minutes. This data release is the combination of over 1,000\,hours of observations taken with the MWA and covers a sky area of $\approx$\survarea{}\,sq.deg. The region of sky covered in this data release and in \dri{} is presented in Figure~\ref{fig:gleamx_sky_coverage}

\section{Continuum pipeline}\label{sec:dr}

The data reduction process used for both this data release and \dri{} is described in full by HW22. However, with the increased sky coverage of this data release, several improvements to the data processing were introduced to improve image quality for low elevation pointings and minimise contamination from bright radio sources. The updated GLEAM-X pipeline is publicly available on the GLEAM-X organisation GitHub\footnote{\url{https://github.com/GLEAM-X/GLEAM-X-pipeline}} and as a containerised pipeline that can be run on any platform with Singularity installed \citep{2017PLoSO..1277459K}. Here we outline the changes to the reduction steps introduced in this data release. As the changes introduced in this data release are typically due to the increased Declination coverage (compared to \dri{} where only zenith pointings were processed), the changes have been integrated into the GLEAM-X pipeline and will be implemented in future GLEAM-X data releases. 

\subsection{Calibration and improved sky model}\label{sec:calibration}

We perform the same calibration approach as \dri{}, by calibrating separately on individual 2-minute snapshot observations in a direction-independent manner. A sky model is used that is primarily derived from GLEAM with additional models for complex sources. In this data release, we updated the model for Pictor-A with an improvement on the amplitude calibration, particularly for short baselines. The model for Pictor-A was derived from a combination of the Very Large Array (VLA) Sky Survey \citep[VLASS;][]{2020PASP..132c5001L} and observations taken with MWA Phase II as part of the GLEAM-X observing. 

Calibration solutions from GLEAM-X observations that were near in time and pointing were applied to the observations of Pictor-A and then imaged using \textsc{WSClean} \citep{2014MNRAS.444..606O}. A multi-component model of Pictor-A was derived with spectral indices calculated using all channels of the MWA and VLASS for each component. The final model comprised four components: two extended lobes with steep spectral indices and two hot spots with flat spectral indices.

\subsection{Sidelobe subtraction}\label{sec:sidelobesub}

For observations taken at a low elevation, the sidelobe of the MWA primary beam has significant sensitivity that can cause imaging artefacts including alias sources from the sidelobe reflected into the main lobe. In this release, we introduce a sidelobe subtraction for observations where the sidelobe was more sensitive than 0.1\% of the primary beam\footnote{The mainlobe and sidelobe locations and sensitivities were calculated using \texttt{get\_mwa\_pb\_lobes.py} as part of the \textsc{skymodel} package available on GitHub: \url{https://github.com/Sunmish/skymodel}}. 

To reduce the sidelobe power, the visibilities are phase rotated to the sidelobe and then a shallow image is formed to create a model of the sidelobe. The shallow image is generated using \textsc{WSCean}, using the following settings: 
\begin{itemize}
    \item imaging the XX, YY, XY, and YX instrumental polarisation with the cleaning performed on peaks calculated from the sum of the squares but subtracted from individual images using the \texttt{-join-polarizations} option;
    \item four 7.68\,MHz channels are jointly cleaned for each polarisation product;
    \item a Briggs robust parameter of $0.5$ \citep{1995AAS...18711202B};
    \item only one major clean cycle, where the images are inverse Fourier transformed back to visibilities and subtracted from the data;
    \item up to $4\times10^{4}$ minor cleaning cycles where subtractions are performed in the image plane rather than the visibilities;
    \item cleaning was stopped when negative components were reached
\end{itemize}

The `MODEL\_DATA' column of the observation measurement set was updated during this imaging, and subtracted from the visibilities after completion. The visibilities were then phase-rotated back to the mainlobe. This procedure is similar to the ``demixing'' often used by LOFAR \citep{2007ITSP...55.4497V,2022A&A...658A...1M}, which also subtracts a bright off-axis nearby source from the data. However, demixing involves calibrating using a model for the source, which is then used to subtract the source from the data. In the sidelobe subtraction procedure introduced here, no calibration is derived or applied, a model of the sensitive sidelobe is produced via imaging and subtracted directly. This procedure dramatically reduced the overall power of the sidelobe sources in the visibilities, particularly for aliased sources (described further in Section~\ref{sec:issues}), and ultimately resulted in a slight decrease to the final image root-mean-square (RMS) noise level. Self-calibration was also applied to any observations that were identified as having significant sidelobes, this is described further in Section~\ref{sec:selfcal}.

\subsection{Self calibration}\label{sec:selfcal}
For observations that were identified as having a significant sidelobe, a round of amplitude and phase self-calibration was applied. There was no noticeable improvement on image quality for observations that had no significant sidelobe contributions, thus self-calibration was not performed on these observations. Visibilities were phase-rotated back to the mainlobe pointing and another shallow imaging procedure was performed using \textsc{WSClean} with the same imaging parameters as the shallow imaging for sidelobe subtraction outlined in Section~\ref{sec:sidelobesub}. The `MODEL\_DATA' column of the observation measurement set was updated as part of the \textsc{WSClean} imaging procedure. The `MODEL\_DATA' column was then used to calibrate using \textsc{mitchcal} \citep{2016MNRAS.458.1057O}. These calibration solutions were only applied if less than 20\% of the solutions were flagged.

\subsection{Ionospheric assessment}\label{sec:io_checks}
A final check of image quality was performed to identify any potential ionospheric effects resulting in significant blurring of sources in the image before combining snapshot images into the final mosaic. The blurring of sources from the ionosphere has a larger impact on the lowest frequencies (i.e. $72$--$103$\,MHz), but the quality check was performed on all frequencies for consistency. 

For each snapshot image, a quality control catalogue of sparse (no nearby sources within 1\,\mbox{\ensuremath{'}}), unresolved in NRAO VLA Sky Syrvey \citep[NVSS;][]{1998AJ....115.1693C} and/or Sydney University Molonglo Sky Survey \citep[SUMSS;][]{2003MNRAS.342.1117M}, and high signal-to-noise (SNR $>=$ 50) sources was produced and used for all further calculations. A source-finding procedure is performed on each snapshot image using \textsc{Aegean}\footnote{\url{https://github.com/PaulHancock/Aegean}} \citep{2012MNRAS.422.1812H, 2018PASA...35...11H}, and a cross-match with the quality control catalogue was performed with 1\,\mbox{\ensuremath{'}} separation. Observations were flagged if there was fewer than 100 sources in the cross-match catalogue, or if the RMS noise level of the image were significantly higher ($3\sigma$) than the average RMS noise level for GLEAM-X images at the corresponding frequency band. The average and standard deviation of the ratio of integrated flux density to peak flux density (${S_{\mathrm{int}}/{S_{\mathrm{peak}}}}$) of sources in the image were calculated for every observation for each night. Observations were flagged as having significant blurring if ${S_{\mathrm{int}}/{S_{\mathrm{peak}}}}$ had a mean $\geq1.1$ or a standard deviation $\geq0.125$. 

\begin{figure}
\begin{subfigure}{\linewidth}
    \centering
    \includegraphics[width=1\textwidth]{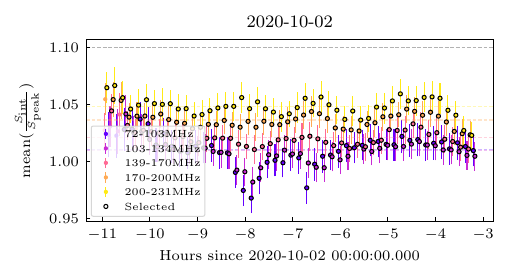}
    \label{fig:good_ionight}
\end{subfigure}
\begin{subfigure}{\linewidth}
    \centering
    \includegraphics[width=1\textwidth]{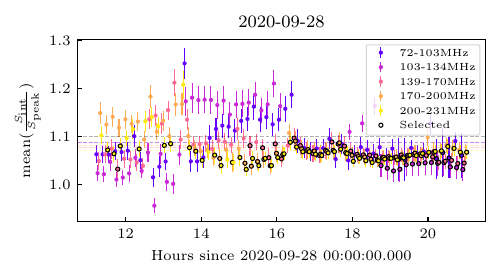}
    \label{fig:bad_ionight}
\end{subfigure}  
\caption{The mean ratio of integrated flux density to peak flux density, ${S_{\mathrm{int}}/{S_{\mathrm{peak}}}}$, for each 2 minute snapshot observation over two nights of observing in this work. The top figure corresponds to 2020-10-02, a night identified as having good ionospheric conditions; and lower figure corresponds to 2020-09-28, a night identified as having poor ionospheric conditions. Horizontal lines of the same colours represent the median value for ${S_{\mathrm{int}}/{S_{\mathrm{peak}}}}$ for the corresponding channel over the night. The horizontal grey dashed line is the limit of 1.1 for ${S_{\mathrm{int}}/{S_{\mathrm{peak}}}}$ above which observations are discarded as having large ionospheric blurring. Points with a black outline are those that were selected as passing the ionospheric analysis and were included in mosaics for this release.}
\label{fig:ionospheric_analysis}
\end{figure}

The mean and standard deviation of ${S_{\mathrm{int}}/{S_{\mathrm{peak}}}}$ were inspected for each observation over each observing night to identify any periods of high ionospheric activity or areas with poor image quality. The lowest frequency band (72--103\,MHz) typically saw the largest number of observations flagged, which is expected since this frequency shows the greatest ionospheric blurring. However, the larger field of view also results in greater overlap between snapshot observations and thus minimal reduction in sensitivity in the final mosaic. This triage typically saw 10--15\% of observations being flagged in the lowest frequency band in a night of observations and <5\% for the higher frequency bands, although up to 50\% of observations were flagged in some nights with particularly large levels of ionospheric activity. Figure~\ref{fig:ionospheric_analysis} presents the mean ${S_{\mathrm{int}}/{S_{\mathrm{peak}}}}$ over a night of good ionospheric conditions and poor ionospheric conditions. 

\subsection{Combined multi-night mosaicking}\label{sec:comb_night}
The mosaicking procedure for individual nights followed the same routine that was outlined in \dri{}. Individual snapshots were weighted according to the square of the primary beam model and the inverse square of the RMS of image. Snapshots were then co-added using \textsc{swarp} \citep{2002ASPC..281..228B} and the point-spread-function (PSF) of the combined mosaic image was measured using the same approach as that outlined in \dri{}. This resulted in a total of 26 mosaic images (20$\times$ 7.68-MHz frequency channel mosaic images, five$\times$ 30.72-MHz mosaic images and a wideband 60-MHz image over 170--231\,MHz) for each night of observations. However, for the Dec~$-71^\circ$ drift scan, multiple nights had at least one frequency band that was of poor quality. Consequently, for the Dec~$-71^\circ$ drift scan observations, instead of four nights each with 26~mosaics, observations from each night were combined in one step into 26~combined four-night mosaics.

The combined final mosaics for this data release were created by combining all 25 declination strips for each of the 20 7.68\,MHz sub-band images and five 30.72\,MHz images using \textsc{swarp}. To optimise the signal-to-noise and ensure a smooth co-addition of the declination strips, weight maps were derived from the inverse square of the RMS maps as measured by \textsc{BANE}, a companion tool of \textsc{Aegean}. Weight maps were also calculated based on a declination-dependent sigmoid function to downweight the edges of the mosaics and minimise artefacts between the different declination strips. The 170--200\,MHz and 200--231\,MHz images were re-gridded and convolved to a common resolution and combined using \textsc{swarp} to create the deep 60\,MHz wideband image used for source-finding.

\subsection{Noise analysis}\label{subsec:noise_analysis}
\begin{figure*}
    \centering
    \includegraphics[width=1\textwidth]{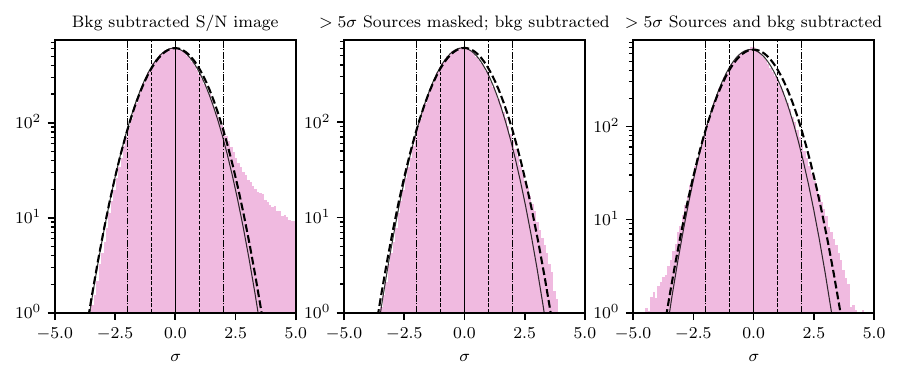}
    \caption{Pixel distribution for a 25 square degrees region of the wideband source-finding image covering 170--231\,MHz. \textsc{BANE} measure an RMS noise level in this region of 0.7\,mJy\,beam$^{-1}$. The left most panel shows the distribution of the S/N of pixels in the image after the background has been subtracted and dividing by the RMS noise map. Sources that are detected at 5$\sigma$ down to 0.2$\sigma$ are then either masked using \textsc{aeres} (central panel) or subtracted (right panel). The black solid lines show Gaussian distributions with $\sigma =1$ (as measured by \textsc{BANE}) and the black dashed Gaussian distribution is the fitted Gaussian to the pixel distribution. A similar distribution as measured by \textsc{BANE} to the pixel distribution indicates confusion is not impacting the effectiveness of \textsc{BANE} at measuring the background and RMS. The vertical solid lines indicate the mean values; dashed lines indicate |S/N|$=1\sigma$; and dash-dotted lines indicate |S/N|$=2\sigma$.  }
    \label{fig:noise_analysis_white}
\end{figure*}

Here we analyse the noise properties of the source finding 170--231\,MHz image and assess the impacts of confusion. We follow the noise analyses of \dri{} and \cite{2017MNRAS.464.1146H} on a 25\,deg$^2$ region centered on RA~$2^\mathrm{h}30^\mathrm{m}$ Dec$-40^\circ00'$ as a representative region with typical source distribution and noise properties. The background and RMS maps are measured using \textsc{BANE} and the background is subtracted from the image. An initial source finding routine using \textsc{Aegean}, is used to detect sources down to $0.2\times$ the local RMS noise level. These sources are either masked or subtracted from the background-subtracted image using \textsc{aeres} from the \textsc{Aegean} package. In Figure~\ref{fig:noise_analysis_white}, we present the histogram of the pixel distribution for the background subtracted, masked and source subtracted images. 

As discussed in HW22, a survey approaching the confusion limit will skew towards a positive distribution. However, for the wideband source finding image, the distribution is almost entirely symmetrical. The higher resolution of GLEAM-X compared to GLEAM means confusion contributes a smaller fraction to the noise \citep{2019PASA...36....4F}. However, at the lowest frequency band of GLEAM-X, 72--103\,MHz, the lower resolution means confusion is contributing a significant fraction to the noise levels. Consequently, \textsc{BANE} is unable to accurately measure the noise levels across the entire mosaic in the lowest frequency band. While the lowest frequency images of GLEAM-X are close to the confusion limit, the higher resolution of the higher frequency images provides sufficient information to reduce the contribution of confusion to the final noise maps. The contribution of confusion to the noise levels at the wideband source finding image is minimal, so we use the source positions from the wideband catalogue for the priorized fitting routine of \textsc{Aegean}. 

\begin{figure*}
    \centering
    \includegraphics[width=1\textwidth]{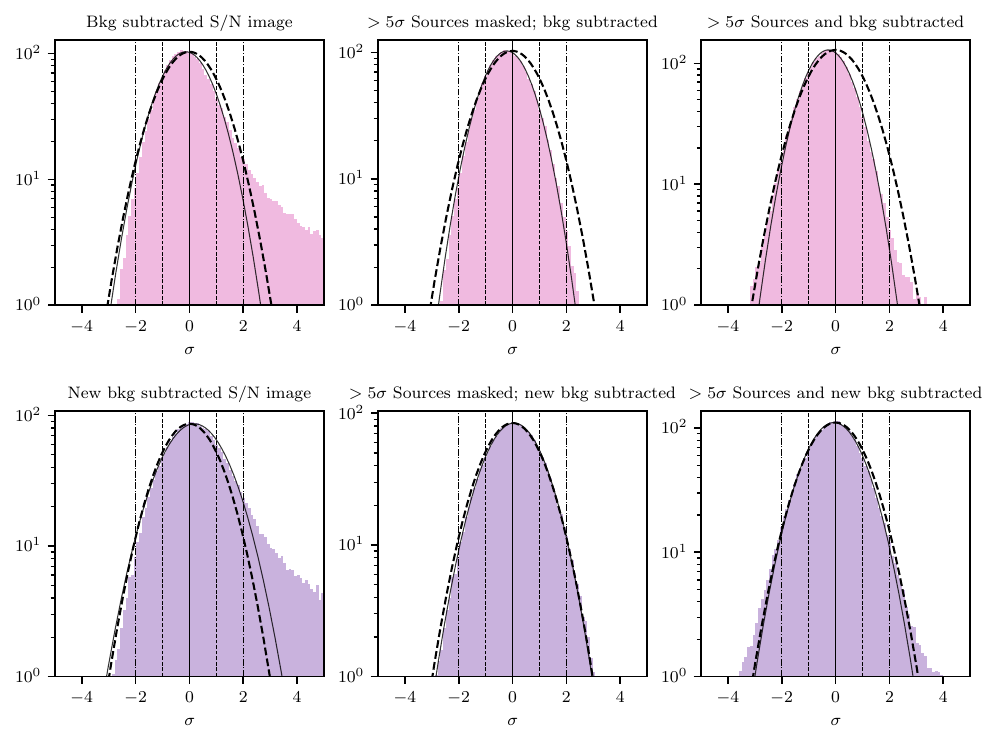}
    \caption{Pixel distribution for a 25 square degrees region of the lowest band image covering 72--103\,MHz. The left, middle and right panels are the same as described in Figure~\ref{fig:noise_analysis_white}. The top three panels use the initial background and RMS maps measured by \textsc{BANE}, while the bottom three panels use updated background and RMS maps measured by \textsc{BANE} after sources that were detected in the wideband source finding image are subtracted. The similarity of the solid line Gaussian distribution (measured by \textsc{BANE}) and the dashed line Gaussian distribution (fit to the pixel distribution) in the bottom three panels shows a dramatic improvement in the background and RMS estimation after sources are subtracted. Likewise, the difference in the distributions in the top three panels, indicates \textsc{BANE} does not accurately measure the background or RMS maps, likely due to confusion. }
    \label{fig:noise_analysis_red}
\end{figure*}

For the 72--103\,MHz image and the four 7.68\,MHz sub-band images, an initial round of source finding using \textsc{Aegean} with the wideband catalogue as a prior for source positions was conducted. This catalogue of sources was subtracted using \textsc{aeres} and the noise and background maps were measured again using \textsc{BANE}. The noise analysis was then performed as described above, to produce a background subtracted image and masked and source subtracted images, using the updated noise and background maps. In Figure~\ref{fig:noise_analysis_red}, we present the histograms of the pixel distributions for the 72--103\,MHz images both before and after the improved background and noise estimates. Using the new background maps, the noise distribution becomes almost completely symmetric and follows the values measured by \textsc{BANE}. 

The increasing resolution with the higher frequencies means only the lowest frequency band, 72--103\,MHz, is significantly impacted by confusion, thus improved noise and background maps were only generated for the 72--103\,MHz image and corresponding 7.68\,MHz sub-band images. All noise and background maps are made available as part of the survey data release. 

\section{Final images}\label{sec:images}
The 26 mosaics produced at the end of the combined multi-night mosaicking described in Section~\ref{sec:comb_night} are the combination of 28 nights of observing in 2020 and are Stokes I images across 72--231\,MHz in five 30.72\,MHz and 20 ``subband'' 7.68\,MHz bands as well as one deep 60\,MHz band image across 170--231\,MHz. There is decreasing sensitivity towards the edges of the mosaic, we therefore select a region with roughly consistent sensitivity covering 20\,h40\,m$\leq$RA$\leq$6\,h40\,m, -90\dg$\leq$Dec$\leq$+30\dg (\survarea{-deg$^2$}) for this release. Postage stamps of all the images of this work are available on both the GLEAM-X website\footnote{\url{https://www.mwatelescope.org/gleam-x}} and AAO Data Central.

\begin{figure*}
    \centering
    \includegraphics[width=\textwidth]{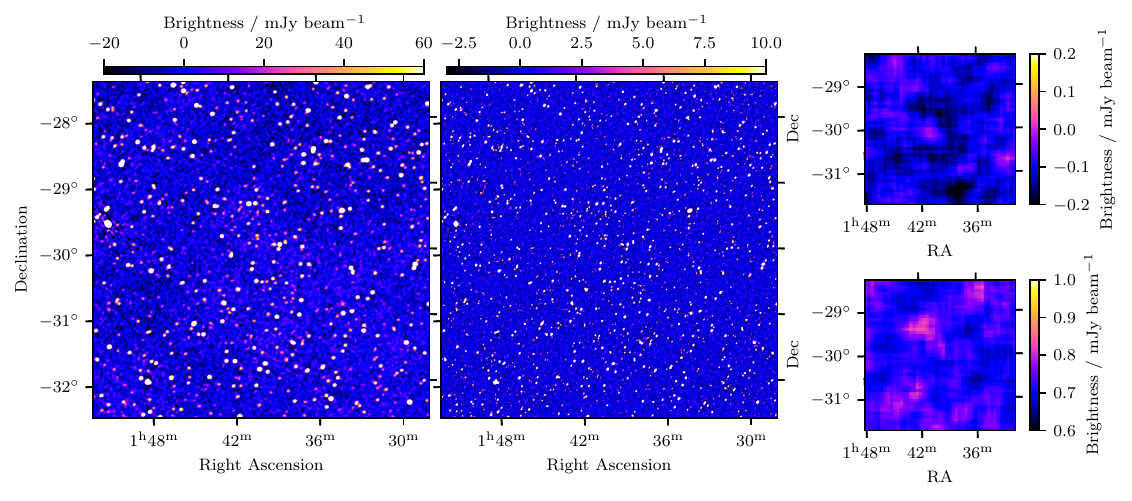}
    \caption{A ten square degree region in GLEAM ExGal and this work centered on 01\,h40\,m RA, -30\dg{} Dec. The left panel shows the region of the source finding 170--231\,MHz mosaic in GLEAM ExGal, the central panel shows the same region in the source finding mosaic of this work, the top and bottom images in the right panel show the corresponding background and RMS noise of the GLEAM-X source finding image in the corresponding region. GLEAM ExGal contains 216 sources in this region, and the average RMS noise is 6\,mJy\,beam$^{-1}$; GLEAM-X contains 811 sources and the average RMS noise level is 0.8\,mJy\,beam$^{-1}$. }
    \label{fig:image_comparisons}
\end{figure*}

As described in Section~\ref{subsec:noise_analysis}, we calculate the noise and background maps for each mosaic using \textsc{BANE} with re-calculated background and RMS maps for the the lowest frequency band and corresponding sub-band images (i.e. 72--103\,MHz). As with \dri{}, we also performed 10 loops of 3-sigma-clipping to exclude the source-filled pixels in the background estimations. We present an example of 10\,sq. deg of the 170--231\,MHz wideband mosaic with its associated background and RMS noise maps and the same region in GLEAM ExGal \citep{2017MNRAS.464.1146H} in Figure~\ref{fig:image_comparisons}. 

The RMS noise of the wideband 170--231\,MHz image has a median of  1.5\,mJy\,beam$^{-1}$, within the expectation of $\sim$1.2\,mJy\,beam$^{-1}$ calculated in \dri{}. Likewise, in the 30\,MHz images we find RMS values ranging from 7--2\,mJy\,beam$^{-1}$, which is also comparable to values reported in \dri{}. While values are slightly higher than those reported in \dri{}, we attribute this to the larger Declination coverage and bright A-team sources included in this data release. Large areas of this data release are found to have RMS noise levels <1\,mJy; we attribute the slight decrease in these areas to the improvements to the data reduction introduced in this release, e.g. the sidelobe subtraction and self calibration.


\section{Compact Source Catalogue}
\label{sec:catalogue}
As with \dri{}, we present a source catalogue derived from the images in this work alongside with the mosaics. The details of the source detection and error derivation of this catalogue are described in full in \dri{}, however, we summarise the detection strategy here. We use the algorithm \textsc{Aegean} to find sources in the most sensitive 170--231\,MHz image that are S/N$>4\sigma$. The positions for all sources in this catalogue are then used for `priorized' fitting on the other images to measure the flux densities using the local PSF of the relevant narrow-band images. During the priorized fitting stage, if \textsc{Aegean} is unable to determine the error for a given parameter, it sets the error to a value of $-1$\footnote{See \url{https://github.com/PaulHancock/Aegean/wiki/AegeanErrors} for details.}, for these reason we use $-1$ as a flag value for the catalogue.

The catalogue is filtered to contain only sources within the defined region of this survey (i.e. 20\,h40\,m$\leq$RA$\leq$6\,h40\,m,  $-90^{\circ}\leq$Dec$\leq+30^{\circ}$) with flux densities $\geq 5\sigma$, where $\sigma$ denotes to local RMS noise. Simple spectral models are fit to all sources in the catalogue and sources that are fit well (described further in Section~\ref{sec:alpha}) have parameters reported in the final catalogue. The spectral modelling procedure is described further in Section~\ref{sec:alpha}.

The final catalogue, after filtering, consists of \nsrc{} radio sources detected over \survarea\,deg$^{2}$ with \nfit{} sources with reported fits for either a power-law or curved spectrum. The increase in source density from GLEAM ExGal to this data release is likely due to a combination of increased resolution and sensitivity. A detailed analysis of source counts of this data release is presented in Venville et al. (submitted).  We present the main statistics of the compact source catalogue for this data release in Table~\ref{tab:survey_stats} as well as a comparison to relevant surveys. The catalogue has \ncol{} columns (described in Appendix~\ref{sec:columns}).

\begin{table*}
\centering
    \caption{Survey properties and statistics GLEAM-X DRII compared to both GLEAM-X DRI and the largest single data release from GLEAM ExGal. Values are given as the mean, $\pm$ the standard deviation where appropriate. The statistics shown are derived from the wideband (170--231\,MHz) image. The internal flux density scale error applies to all frequencies. \label{tab:survey_stats}}
    \begin{tabular}{cccc}
    \hline
    Property & GLEAM-X~DRII & GLEAM ExGal & GLEAM-X~DRI \\
    \hline
    Number of sources & \nsrc{} & 307,456 & 78,967 \\
    Number of sources spectrally fit & \nfit{} & 254,453 & 71,320 \\
    Sky area & \survarea{}~deg$^{2}$ & 24,402~deg$^{2}$ & 1,447~deg$^{2}$ \\
    Source density & 48~deg$^{-2}$ & 13~deg$^{-2}$ & 55~deg$^{-2}$ \\
    RA astrometric offset & $ -7\pm800$\,mas & $-4\pm16''$ & $+14\pm700$\,mas \\
    Dec astrometric offset  & $ +4\pm800$\,mas & $0.1\pm3.6''$ & $+21\pm687$\,mas \\
    Internal flux density scale error & 2\,\% & 2\,\% & 2\,\% \\
    50\,\% completeness & 5.8\,mJy & 55\,mJy & 5.6\,mJy \\
    90\,\% completeness & 10.2\,mJy & 170\,mJy & 10\,mJy  \\ 
    98\,\% completeness & 50\,mJy & 500\,mJy  & 50\,mJy \\ 
    Reliability for $S_\mathrm{int}\geq7\sigma$ & \pctreliablehigh{}\,\% & 99.8\,\%  & \pctreliablehigh{}\,\% \\
    Reliability for $S_\mathrm{int}\geq5\sigma$ & \pctreliablelow{}\,\% & 98.9\,\% &  \pctreliablelow{}\,\% \\
    Image RMS noise & $1.5^{+1.5}_{-0.5}$\,mJy beam$^{-1}$ & $11.3\pm7.3$\,mJy\,beam$^{-1}$  & $1.27\pm0.15$\,mJy\,beam$^{-1}$  \\
    PSF major axis & $85\pm18''$ & $152\pm25''$ & $77\pm12''$ \\
    PSF minor axis & $64\pm8''$ & $134\pm12''$ & $61\pm6''$ \\
    \hline
    \end{tabular}
\end{table*}

\subsection{Comparison with DRI and GLEAM}\label{sec:gleam}
Both \dri{} and this data release use GLEAM as the basis for flux density calibration. Here we compare the flux densities measured in this work with GLEAM ExGal. A catalogue of sources that are compact in both catalogues ($S_{\mathrm{int}}/S_{\mathrm{peak}}<2$), cross-match within a 15$''$ radius, and have a confident power-law spectral index fit (reduced $\chi ^2<1.93$, corresponding to a 99\% confidence level) is used for all the following analysis. 

\begin{figure}
    \centering
    \includegraphics[width=\linewidth]{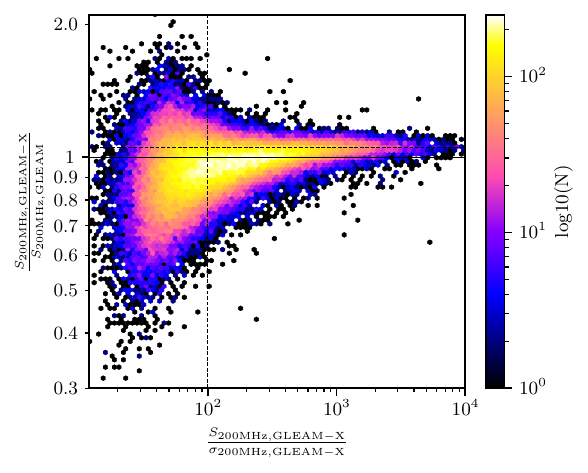}
    \caption{Ratio of the 200\,MHz integrated flux density for compact sources matched in GLEAM-X DRII and GLEAM as a function of signal-to-noise in GLEAM-X. The vertical dashed line is at a signal-to-noise of 100, corresponding to roughly 90\% completeness in GLEAM. The horizontal solid line corresponds to a ratio of 1, and the horizontal dashed line corresponds to a ratio of 1.05, which fits the trend better. The same trend was detected in \dri{}. Colour represented a density of points, error bars are omitted for clarity but are calculated as the quadrature sum of the measurement errors in both surveys.}
    \label{fig:flux_ratio_gxvg}
\end{figure}

We follow the same analysis outlined in \dri{} and compare the ratio integrated flux densities as a function of signal to noise for both \drii{} and GLEAM ExGal. The comparison of these ratios is presented in Figure~\ref{fig:flux_ratio_gxvg}. As with \dri{}, there is a trend towards 1.05 at higher signal to noise sources, however, as the effect is minimal we do not correct for it here.  We advise an 8\% error for the flux density scale of \drii{} when comparing to other surveys, based on the 8\% relative error of GLEAM for which our flux density scale is based. 

\begin{figure}
    \centering
    \includegraphics[width=\linewidth]{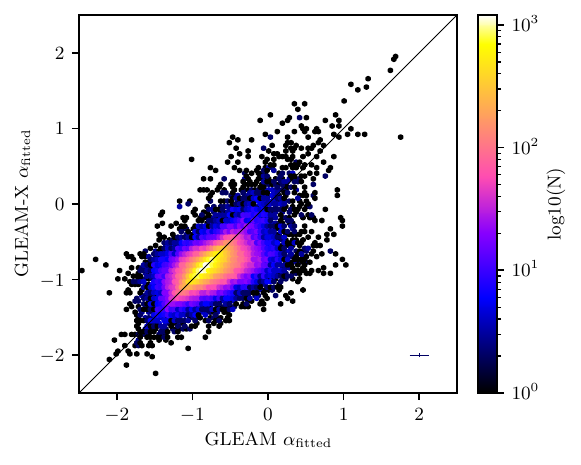}
    \caption{Spectral indices, $\alpha$, based on a power-law spectral model, across the 7.68\,MHz narrow bands for compact sources matched in both GLEAM and GLEAM-X. Colour corresponds to the density of points and an average of the error bars for fitting errors is shown at the bottom right, the fitting errors for GLEAM-X are significantly smaller due to the increase in signal to noise and thus confidence in spectral fitting for a given source. The diagonal line shows a 1:1 ratio of $\alpha$. }
    \label{fig:alpha_comparison}
\end{figure}

We also compare the fitted spectral indices for components best fit by a power-law spectral model ($\alpha_{\mathrm{fitted}}$), presented in Figure~\ref{fig:alpha_comparison}, and find no clear trends. The increase in signal to noise of GLEAM-X does result in consistently smaller error bars for $\alpha_{\mathrm{fitted}}$.

\subsection{Spectral fitting}\label{sec:alpha}
We fit two spectral models to the 20 narrow-band flux density measurements for all detected sources in the filtered compact source catalogue described in Section~\ref{sec:catalogue}, we define the spectral index, $\alpha$, as $S\propto \nu^{\alpha}$, such that a negative $\alpha$ describes a negative slope in logarithmic space. 

\begin{figure}
\begin{subfigure}{\linewidth}
    \centering
    \includegraphics[width=1\textwidth]{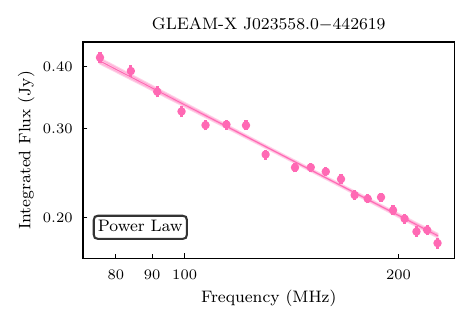}
    \label{fig:sed_pl}
\end{subfigure}
\begin{subfigure}{\linewidth}
    \centering
    \includegraphics[width=1\textwidth]{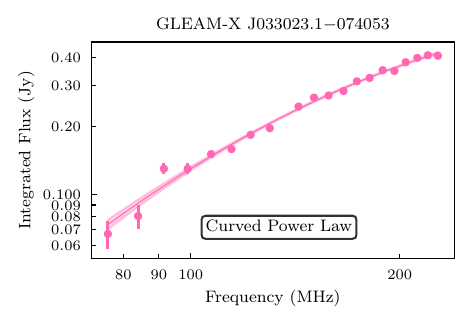}
    \label{fig:sed_curved}
\end{subfigure}  
\begin{subfigure}{\linewidth}
    \centering
    \includegraphics[width=1\textwidth]{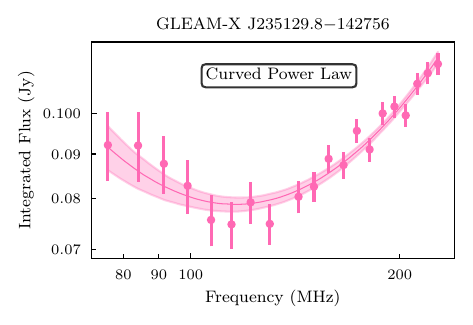}
    \label{fig:sed_uptick}
\end{subfigure}  
\caption{Example SEDs for 3 sources included in this data release. The title includes the source name as included in the catalogue, the inset includes the spectral model identified as the best fit model as outlined in Section~\ref{sec:alpha}. The optimised model and its 1$\sigma$ confidence interval is overlaid as the pink line and shaded region of each source. The `Power Law' and `Curved Power Law' models are as defined by Equation~\ref{eq:plaw} and Equation~\ref{eq:curve_model} respectively.}
\label{fig:seds}
\end{figure}

Sources are first fit with a simple power-law model parameterised as: 

\begin{equation}
    S_{\nu} = S_{\nu_0}\left( \frac{\nu}{\nu_0}\right)^{\alpha},
    \label{eq:plaw}
\end{equation}

where $S_{\nu_0}$, is the flux density in Jy at the reference frequency, $\nu_0$. The large fractional bandwidth of MWA allows for curvature to be detected and characterised within the MWA bandwidth. We therefore also fit a modified power-law model that parameterises the spectral curvature: 

\begin{equation}
    \label{eq:curve_model}
    S_{\nu} = S_{\nu_0}\left( \frac{\nu}{\nu_0}\right)^{\alpha}\exp{\left(q\ln\left({\frac{\nu}{\nu_0}}\right)^2\right)},
\end{equation}
where increasing $|q|$ describes increasing curvature, $q<0$ corresponds to a concave curve and $q>0$ corresponds to a convex curve \citep{2012MNRAS.421..108D}. Equation~\ref{eq:curve_model} has no physical motivation and is used as an initial identification of potential peaked-spectrum sources (PSS). A comprehensive catalogue of PSS in GLEAM-X will be given in a separate publication (Ross et al. in prep). In Figure~\ref{fig:seds}, shows example SEDs for three sources in this release: a source best fit with a typical, linear power-law, and two best fit with a curved power-laws. 

Both the simple power-law and curved power-law models are fit using the \textsc{SciPy} \textsc{python} module that applies a Levenberg-Marquardt non-linear least-squares regression algorithm \citep{2020SciPy-NMeth}. Narrow-band measurements that had a negative integrated flux density were excluded from fitting and fits are reported for sources if both of the following are true: 

\begin{itemize}
    \item there were at least 15 integrated flux density measurements;
    \item the $\chi^2$ goodness-of-fit was above 99\% likelihood confidence.
\end{itemize}

Similarly, a curved model is reported instead of the power-law model if the following criteria were all met: 
\begin{itemize}
    \item $q/\Delta q \geq 3$;
    \item $|q| > 0.2 $;
    \item the reduced $\chi^2$ for a power-law model was higher than the reduced $\chi^2$ for a curved power-law model.      
\end{itemize}
By adding the criteria $|q|>0.2$, we avoid the potential for favouring a curved model over a power-law model for spectra that show only a small level of curvature \citep{2017ApJ...836..174C}. 

As with \dri{}, the internal flux density scaling consistency of the catalogue is verified using the reduced $\chi^2$ of the model fitting. We adopt 2\% as the flux density scaling error as this produces a consistent median reduced $\chi^2$ of unity as a function of signal to noise.

\begin{figure}
    \centering
    \includegraphics[width=\linewidth]{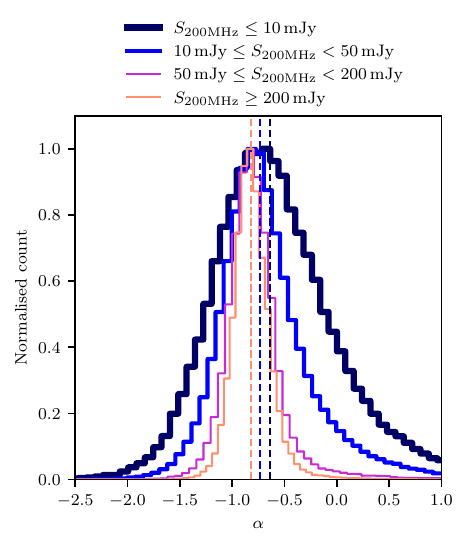}
    \caption{Distributions of the spectral index, $\alpha$, for sources where the fit was successful, for various flux density bins. The dark navy line shows sources with $S_\mathrm{200MHz}<10$\,mJy, the blue shows sources with $10\leq S_\mathrm{200MHz}<50$\,mJy, the purple line shows sources with $50\leq S_\mathrm{200MHz}<200$\,mJy, and the orange line shows sources with $S_\mathrm{200MHz}>200$\,mJy. The dashed vertical lines of the same colours show the median values for each flux density cut: $-0.58$, $-0.77$, $-0.84$, and $-0.84$, respectively.}
    \label{fig:alpha_dist}
\end{figure}

The distributions of $\alpha$ as a function of flux densities are presented in Figure~\ref{fig:alpha_dist}. The reported spectral indices of this data release are consistent with both GLEAM and \dri{}, with a median $\alpha$ in the brightest flux density bin of $-0.84$. 

\subsection{Astrometry}\label{sec:overall_astrometry}
Following \citet{2017MNRAS.464.1146H} and HW22, the astrometry is calculated using the wideband 170--231\,MHz reference catalogue. Only GLEAM-X sources with a high signal-to-noise ($\geq50\sigma$) were used to calculate offsets, corresponding to a total of 107,323 sources used for the astrometric analysis. A reference catalogue combining both SUMSS and NVSS was generated by filtering to include only sparse (no internal cross matches within 3$'$) and unresolved ($S_{\mathrm{int}}/S_{\mathrm{peak}}<1.2$) sources. For astrometric calculations, the reported positions in the reference catalogue were assumed correct and offsets calculated relative to those positions. We find an average RA astrometric offset of $ -7\pm800$\,mas and Declination astrometric offset of $ +4\pm800$\,mas, where errors on the astrometry are calculated from 1 standard deviation. In the final source catalogue presented in this work, we also report fitting errors on the positions for all sources, which typically are larger than the average astrometric offsets calculated here, we therefore do not correct for these offsets.

\begin{figure}
    \centering
    \includegraphics[width=\linewidth]{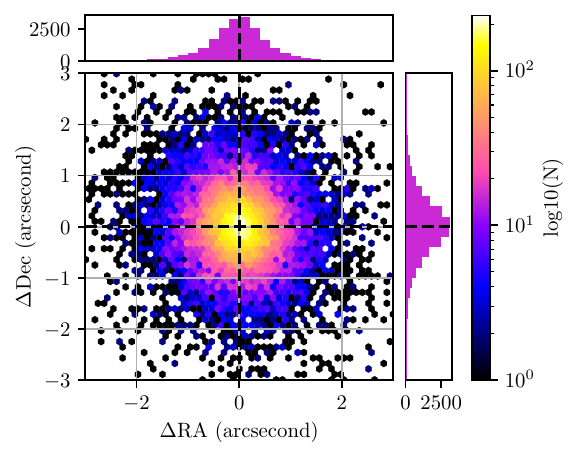}
    \caption{The astrometric offsets of 107,323 isolated, compact, $>50$-$\sigma$ sources after cross-matching against the NVSS and SUMSS reference catalogue described in \sect~\ref{sec:overall_astrometry}. Colour denotes density of points on a log scale. Vertical and horizontal dashed lines indicate the mean offset values in the RA and Dec directions, respectively. Similarly, the horizontal and vertical histograms highlight the counts of the astrometry offsets in each direction.}
    \label{fig:astrometry}
\end{figure}

It is worth noting, during the processing of individual snapshot images, we conduct an astrometric offset correction based on positions reported in the same reference catalogue based on SUMSS and NVSS. For this reason, while we report sub-arcsecond astrometric offsets in the final mosaic, this is not necessarily a true representation of the astrometric properties of this catalogue. With more widefield low frequency surveys of the Southern hemisphere being released, future analyses comparing the astrometric positions reported in this catalogue with an independent radio catalogue may be possible. We suggest care is taken when cross-matching this catalogue with other surveys, particularly at higher frequencies. We present the density distribution of the astrometric offsets in Figure~\ref{fig:astrometry}.




\subsection{Completeness}
\label{sec:completeness}
We follow the same procedure as \citet{2017MNRAS.464.1146H} and HW22, and use the wideband source finding mosaic to simulate completeness of the source catalogue. We inject 130,000 simulated point sources evenly distributed across the region of this release, into the 170--231\,MHz wideband mosaic. The simulated sources are injected with 26 realisations for different flux density increments spanning $10^{-3}$ to ${10^{-0.5}}$\,Jy. Positions for all 90,000 simulated sources remain constant for each realisation, and to avoid confusion, simulated sources are separated by at least 5$'$. The shape of the simulated sources is simulated based on the local major and minor axis of the PSF, and injected into the mosaic using \textsc{aeres}. 

The same source finding procedure as described in Section~\ref{sec:catalogue} is performed on the mosaic with injected simulated sources for each of the 26 realisations. We then calculate the fraction of simulated sources that are recovered to estimate the completeness. For any simulated source that was detected but was near a real source, it was only included as ``recovered'' if the recovered source position was closer to the simulated source rather than the real source. 

\begin{figure*}
    \centering
    \includegraphics[width=\textwidth]{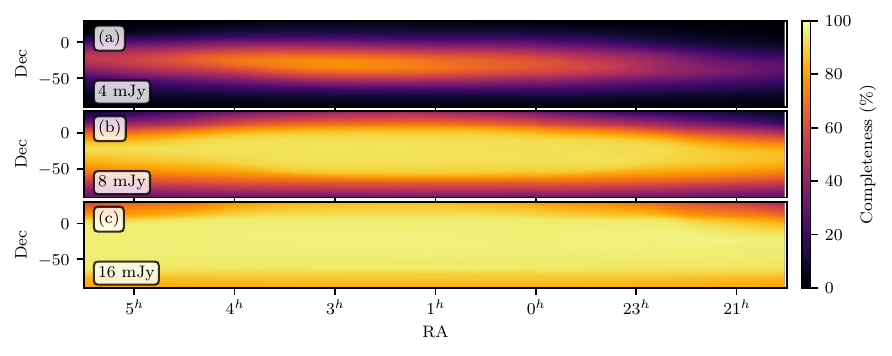}
    \caption{Completeness of the compact source catalogue of this work as a function of sky position for three representative cuts in source integrated flux density at 200\,MHz. The three flux density cuts correspond to completeness levels of approximately 20\%, 75\% and 95\% shown in the subplots a), b) and c) respectively. }
    \label{fig:completeness_spatial}
\end{figure*}

\begin{figure}
    
\centering
\includegraphics[width=1\linewidth]{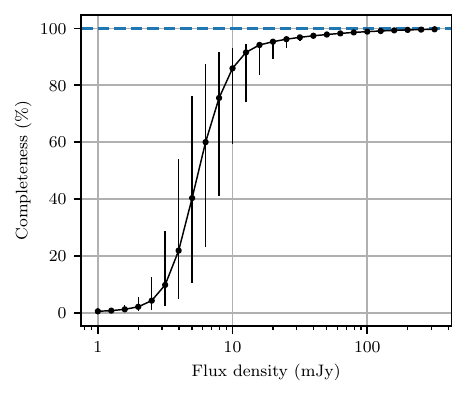}
\caption{GLEAM-X DRII completeness as a function of flux density in the wide-band source finding mosaic covering 170--231\,MHz. The RMS noise is roughly $1.5^{+1.5}_{-0.5}$\,mJy beam$^{-1}$. Larger vertical bars compared to \dri{} are due to the variations in completeness at high and low declinations and in regions near bright A-team sources. }
\label{fig:completeness_curve}
\end{figure}

In \dri{}, the completeness was found to be around 50\% at $\sim$5.6\,mJy and 90\% at $\sim$10\,mJy. In this release, we find completeness that is overall consistent with expectations. The completeness for this work is estimated to be 50\% at $\sim$5.8\,mJy and 90\% at $\sim$10.2\,mJy. Figure~\ref{fig:completeness_spatial} presents the spatial distribution of fraction of simulated sources recovered. The smooth mosaicking of multiple drift scans in this release has produced a near uniform sensitivity and completeness across the region in this release, as expected. However, there is some dependence on RA due to bright A-team sources (most notably, Cygnus A and the Crab nebula) at RA$\approx$5\,h Dec$\approx$+20\,d and RA$\approx$21\,h Dec$\approx$+20\,d. There is a small Declination dependence for high and low Declinations. For high Declination (Dec > 0), observations are taken at low elevation, looking through a larger amount of ionosphere, and typically have more data flagged for poor data quality issues. Consequently, there is a roll off in completeness at high Declination (Dec >0). However, the roll off in completeness at low Declination (Dec <-80) is more likely due to issues with recovering the simulated sources and differentiating from real sources due to the small sky area around the South Galactic Pole. Figure~\ref{fig:completeness_curve} shows the fraction of simulated sources recovered as a function of the flux density at 200\,MHz. The variations in the completeness results in the large error bars in Figure~\ref{fig:completeness_curve}. 

\subsection{Reliability and known issues}
\label{sec:issues}
Following the reliability analysis of \dri{} to check how many false detections may be present, we perform a source finding procedure to identify only negative peaks. Initially, we identify 10,305 negative detections, with 1,316 detections with $S_{\mathrm{peak}}>5\sigma$. As identified in \dri{}, there is a tendency for artefacts around bright positive sources producing both negative and positive detections. The filter identified in \dri{} was applied to exclude both the positive and negative detections that were near these bright sources. Furthermore, a second filter was applied to the negative sources to exclude any that were within 2$'$ of a positive source, as these are likely a result of faint sidelobes of sources that were not properly cleaned. 

\begin{figure}
    \centering
    \includegraphics[width=\linewidth]{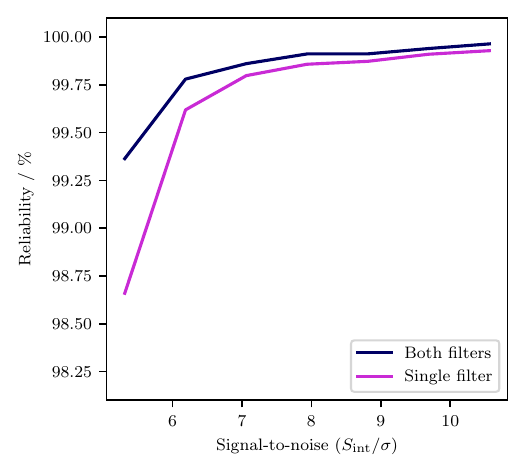}
    \caption{Estimates of the reliability of the catalogue as a function of signal-to-noise. The lower purple curve is a conservative estimate before filtering sources near bright positive sources. The upper dark navy line is a derived after these sources have been filtered out. GLEAM ExGal has a reliability of $\sim$98.9\%-99.8\% at these signal-to-noise levels. }
    \label{fig:reliability}
\end{figure}

We compare the negative and positive detections after applying the first filter and the second filter as a function of signal-to-noise. The reliability for each significance bin is presented in Figure~\ref{fig:reliability}. For a conservative lower limit on the reliability for signal-to-noise at 5$\sigma$, we do not apply the second filter. We find the number of false detections to be 1.3\% which falls quickly to under 1\% at 7$\sigma$ and a plateau for signal-to-noise $>7\sigma$. 

\begin{figure*}
    \centering
    \includegraphics[width=\textwidth]{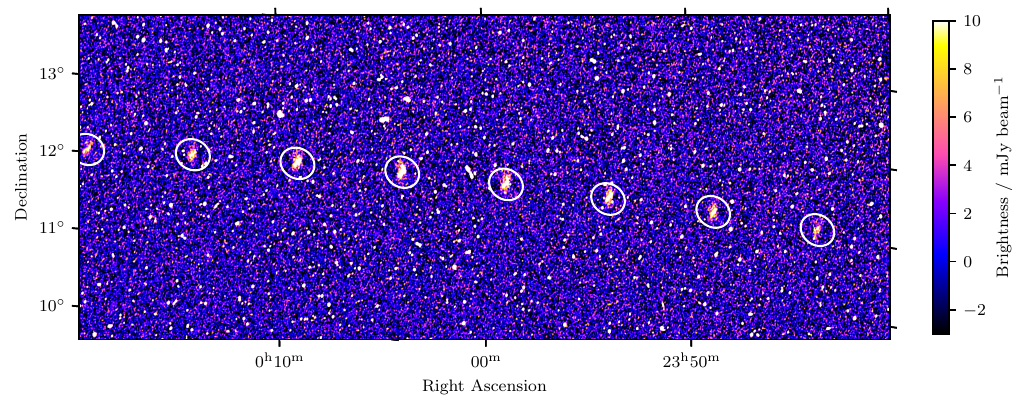}
    \caption{A cutout of the region with an identified repeating artefact due to bright contaminating sources in the sidelobe. The repeating artefact is highlighted by the white circles. }
    \label{fig:phantom}
\end{figure*}

In this analysis, a region was identified as containing a higher density of high signal-to-noise negative detections relative to the rest of the image. This region, covering roughly 23\,h$\leq$RA$\leq$1\,h, +15\dg{}$\leq$Dec$\leq$+30\dg{}, largely overlaps with a region excluded from GLEAM ExGal due to poor ionospheric conditions during observations. It is possible the combination of the low elevation pointing with sparse to no coverage of GLEAM used for calibration in this processing is contributing to the slightly lower reliability in this region. Furthermore, due to a bright source in the sidelobe (PKS 2365--61), for a subset of the observations taken for the +20 Declination drift scans, there is an artefact that appears as a bright source repeating at regular intervals across the high frequency mosaics (170--200\,MHz, 200--231\,MHz, and the sub-bands for each wide images as well as the deep source finding mosaic 170--231\,MHz). The sidelobe subtraction procedure, described in Section~\ref{sec:sidelobesub}, reduces the flux density of this phantom source, but does not remove the artefact entirely. The repeating artefact in this region is presented in Figure~\ref{fig:phantom}. 

\section{Summary and Outlook} \label{sec:summary}

In this work, we present the second data release of the GLEAM-X survey, comprising of images and an extragalactic source catalogue. This data release covers \survarea{}-deg$^2$ covering 20\,h40\,m$\leq$RA$\leq$6\,h40\,m, -90\dg{}$\leq$Dec$\leq$+30\dg{} over 72--231\,MHz in the form of 26 mosaics with bandwidths 60, 30 and 8\,MHz. These mosaics have typical RMS noise levels ranging from $\approx1$\,mJy\,beam$^{-1}$ in the 60\,MHz bandwidth mosaic used for initial source finding to $\approx$10\,mJy\,beam$^{-1}$ in the 8\,MHz mosaic at the lowest frequency, 72--80\,MHz. The RMS noise levels of this region are an order of magnitude lower than that reported in GLEAM ExGal \citep{2017MNRAS.464.1146H} and within expectations for the overall survey as reported by HW22. 

We also present a catalogue of \nsrc{} components, \nfit{} of which are well fit by either a power-law or curved power law spectral model. This results in a source density of \srcdensity{}\,deg$^{-2}$. We estimate this catalogue is 98\% complete at 50\,mJy for the entire region with near uniform sensitivity and completeness across the region covered in this release, as expected. We estimate a reliability of 98.7\% at a 5$\sigma$ level. We identify \ncplfit{} sources as being better fit by a curved power-law, suggesting an order of magnitude increase in the PSS sources from those identified using GLEAM \citep{2017ApJ...836..174C}, and probing a population of fainter PSS. A comprehensive catalogue of PSS and population analysis will be described in an upcoming paper by Ross et al. (in preparation). 
 
The data reduction of this release largely follows that of GLEAM-X~DRI \citep{2022PASA...39...35H}, but with some noticeable improvements that are now incorporated to future GLEAM-X processing. The subtraction of a model of sensitive sidelobes for observations taken at low elevation reduced the occurrence and/or S/N of alias sources appearing in the main lobe. The use of self-calibration for fields with bright and complex sources in the main lobe or sensitive sidelobes was also introduced and resulted in an improvement in the RMS noise levels of snapshot images. Lastly, a stricter quality assessment of blurring due to the ionosphere resulted in fewer poor quality snapshot observations being included in the final mosaics of this release, resulting in an overall improvement in the mosaic quality. 

The uniform and high source density provided by this data release has enabled a study of the source count statistics and a measurement of the angular correlation function, which are presented by Venville et al. (submitted). A search for transients on seconds to hours timescales has also been carried out and will be described in full by Horvath et al. (in prep). 

We aim to release the remainder of GLEAM-X via a series of releases. The POlarised GLEAM-X Survey (POGS-X) will be described fully by Zhang et al. (in prep); a Galactic plane release combining both GLEAM and GLEAM-X observations using joint deconvolution, which will be described in Mantovanini et al. (in prep); finally, we aim to produce a contiguous all-sky coverage and release the remainder of GLEAM-X (Ross et al. in prep). 

The images and source catalogue of the GLEAM-X~DRII data release are publicly available at AAO Data Central (\url{https://datacentral.org.au/services/cutout/}).

\begin{acknowledgements}

NHW is supported by an Australian Research Council Future Fellowship (project number FT190100231) funded by the Australian Government. This scientific work uses data obtained from Inyarrimanha Ilgari Bundara / the Murchison Radio-astronomy Observatory. We acknowledge the Wajarri Yamaji People as the Traditional Owners and native title holders of the Observatory site. Establishment of CSIRO's Murchison Radio-astronomy Observatory is an initiative of the Australian Government, with support from the Government of Western Australia and the Science and Industry Endowment Fund. Support for the operation of the MWA is provided by the Australian Government (NCRIS), under a contract to Curtin University administered by Astronomy Australia Limited. This work was supported by resources provided by the Pawsey Supercomputing Research Centre with funding from the Australian Government and the Government of Western Australia.
Access to Pawsey Data Storage Services is governed by a Data Storage and Management Policy (DSMP). ASVO has received funding from the Australian Commonwealth Government through the National eResearch Collaboration Tools and Resources (NeCTAR) Project, the Australian National Data Service (ANDS), and the National Collaborative Research Infrastructure Strategy. This paper makes use of services or code that have been provided by AAO Data Central (datacentral.org.au). This research has made use of NASA’s Astrophysics Data System Bibliographic Services. The following software was used in this work:
{\sc aoflagger} and {\sc cotter} \citep{2012A+A...539A..95O}; \textsc{WSClean} \citep{2014MNRAS.444..606O,2017MNRAS.471..301O}; {\sc Aegean} \citep{2018PASA...35...11H}; {\sc miriad} \citep{Miriad}; \textsc{NumPy} \citep{NumPy,harris2020array}; \textsc{AstroPy} \citep{Astropy}; \textsc{SciPy} \citep{SciPy}; \textsc{Matplotlib} \citep{Matplotlib}. We also made extensive use of the visualisation and analysis packages DS9\footnote{\href{ds9.si.edu}{http://ds9.si.edu/site/Home.html}} and Topcat \citep{Topcat}. This work was compiled in the very useful online \LaTeX{} editor Overleaf.

\end{acknowledgements}

\begin{appendix}

\setcounter{table}{0}
\renewcommand{\thetable}{A\arabic{table}}

\section{OBSERVATIONS}
\topcaption{GLEAM-X DRII observing summary of the 28 nights published in this work. The HA and Dec are fixed to the locations shown and the sky drifts past for the observing time shown. Observations typically start just after sunset and stop just before sunrise. Nights identified as having high ionospheric activity by \citet{2022PASA...39...35H} are marked with a ``*''. \label{tab:obs}}
\tablefirsthead{\toprule Date & HA & Dec ($\arcdeg$) & Observing time (hours) \\ \midrule}
\tablehead{\multicolumn{4}{c}{{\bfseries  Continued from previous column}} \\ \toprule 
Date & HA & Dec ($\arcdeg$) & Observing time (hours) \\ \midrule}
\tabletail{ \midrule \multicolumn{4}{c}{{Continued on next column}} \\ \midrule}
\tablelasttail{\bottomrule}
\begin{supertabular}{cccc}
2020-09-28* & $-1$ & $-71$ & 9.8 \\
2020-09-29 & $-1$ & $-54$ & 9.3 \\
2020-09-30* & $-1$ & $-39$ & 8.2 \\
2020-10-01* & $-1$ & $-25$ & 7.8 \\
2020-10-02 & $-1$ & $-11$ & 7.9 \\
2020-10-03 & $-1$ & $+2$ & 9.8 \\
2020-10-04* & $-1$ & $+19$ & 9.8 \\
2020-10-05 & 0 & $-72$ & 9.8 \\
2020-10-06 & 0 & $-55$ & 9.8 \\
2020-10-07 & 0 & $-40$ & 9.8 \\
2020-10-08 & 0 & $-26$ & 9.8 \\
2020-10-09 & 0 & $-12$ & 9.6 \\
2020-10-10* & 0 & $+1$ & 8.8 \\
2020-10-11 & 0 & $+18$ & 9.8 \\
2020-10-12 & 0 & $-72$ & 8.6 \\
2020-10-13 & 0 & $-55$ & 8.4 \\
2020-10-14 & 0 & $-40$ & 5.6 \\
2020-10-15 & 0 & $-26$ & 9.1 \\
2020-10-16 & 0 & $-12$ & 9.0 \\
2020-10-17* & 0 & $+1$ & 9.8 \\
2020-10-18 & 0 & $+18$ & 9.7 \\
2020-10-19 & $+1$ & $-71$ & 9.5 \\
2020-10-20 & $+1$ & $-54$ & 9.5 \\
2020-10-21 & $+1$ & $-39$ & 9.5 \\
2020-10-22 & $+1$ & $-25$ & 8.2 \\
2020-10-23 & $+1$ & $-11$ & 9.5 \\
2020-10-24* & $+1$ & $+2$ & 9.5 \\
2020-10-25 & $+1$ & $+19$ & 8.0 \\
\hline
& &  \multicolumn{1}{r}{\textbf{Total:}} & 253.9 \\
\end{supertabular}

\section{Noise analysis}\label{sec:noise_worstcase}
\begin{figure*}
    \centering
    \includegraphics[width=1\textwidth]{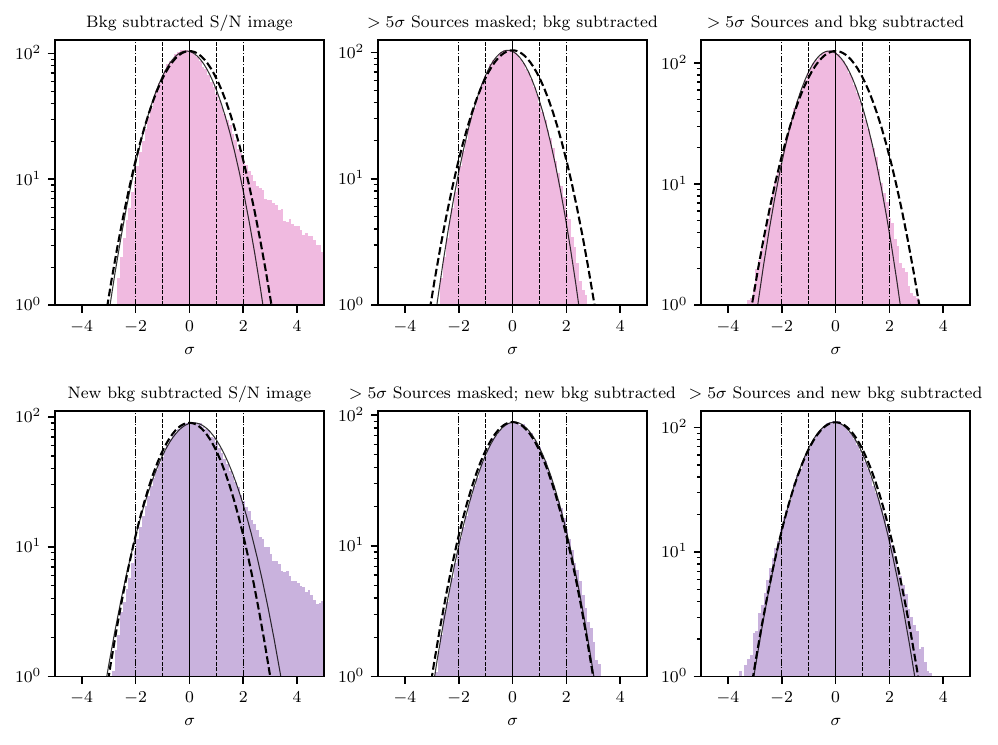}
    \caption{Pixel distribution for a 25 square degrees region of the lowest band image covering 72--103\,MHz centered at RA~$2^\mathrm{h}30^\mathrm{m}$  Dec$+15^\circ00'$. All six panels are the same as described in Figure~\ref{fig:noise_analysis_red}: The top three panels use the initial background and RMS maps measured by \textsc{BANE}, while the bottom three panels use updated background and RMS maps measured by \textsc{BANE} after sources that were detected in the wideband source finding image are subtracted. The similarity of the solid line Gaussian distribution (measured by \textsc{BANE}) and the dashed line Gaussian distribution (fit to the pixel distribution) in the bottom three panels shows a dramatic improvement in the background and RMS estimation after sources are subtracted. Likewise, the difference in the distributions in the top three panels, indicates \textsc{BANE} does not accurately measure the background or RMS maps, likely due to confusion. }
    \label{fig:noise_analysis_red_bad}
\end{figure*}

Given the large sky coverage of this data release, we repeat the noise analysis outlined in Section~\ref{subsec:noise_analysis} for a region of lower quality and lower elevations for the MWA. In Figure~\ref{fig:noise_analysis_red_bad}, we present the pixel distributions for a 25 square degrees region in the lowest wide-band image covering 72--103\,MHz centered at RA~$2^\mathrm{h}30^\mathrm{m}$ Dec$+15^\circ00'$. The lower elevations of this region can elongate the PSF, however, in \dri{}, we outlined a strategy for ensuring the PSF is well defined over the entire region. Consequently, we find the well defined PSF, even in areas of low elevation, ensures we are able to accurately detect and subtract sources from the lowest band in order to reduce the impact of confusion. Comparing the distributions for the optimal scenario (presented in Figure~\ref{fig:noise_analysis_red}), to that of the ``worst case'' scenario in Figure~\ref{fig:noise_analysis_red_bad}, we can see the strategy for estimating the noise and RMS maps is still appropriate and we are not significantly impacted by confusion. 

\section{Catalogue Column Names}\label{sec:columns}

\onecolumn
\topcaption{Column numbers, names, and units for the catalogue.
Source names follow International Astronomical Union naming conventions for co-ordinate-based naming.
Background and RMS measurements were performed by \textsc{BANE} (\Sect~\ref{sec:images});
PSF measurements were performed using in-house software as described in \citet{2022PASA...39...35H};
the fitted spectral index parameters were derived as described in \Sect~\ref{sec:alpha};
all other measurements were made using \textsc{Aegean}. \textsc{Aegean} incorporates a constrained fitting algorithm.
Shape parameters with an error of $-1$ indicate that the reported value is equal to either the upper or lower fitting constraint.
The columns with the subscript ``wide'' are derived from the 200\,MHz wide-band image.
Subsequently, the subscript indicates the central frequency of the measurement, in MHz.
These sub-band measurements are made using the priorised fitting mode of Aegean, where the position and shape of the source are determined from the wide-band image, and only the flux density is fitted (see \Sect~\ref{sec:catalogue}).
Note therefore that some columns in the priorised fit do not have error bars, because they are linearly propagated from the wideband image values (e.g. major axis $a$). \label{tab:catalogue}}
\tablefirsthead{\toprule Number & Name & Unit & Description \\ \midrule}
\tablehead{\multicolumn{4}{c}{{\bfseries  Continued from previous page}} \\ \toprule 
Number & Name & Unit & Description \\ \midrule}
\tabletail{ \midrule \multicolumn{4}{c}{{Continued on next page}} \\ \midrule}
\tablelasttail{\bottomrule}
\begin{supertabular}{llll}
1 & Name & hh:mm:ss+dd:mm:ss & International Astronomical Union name  \\ 
2 & background\_wide & Jy\,beam$^{-1}$ & Background in wideband image \\ 
3 & local\_rms\_wide & Jy\,beam$^{-1}$ & Local RMS in wideband image \\ 
4 & ra\_str & hh:mm:ss & Right ascension \\ 
5 & dec\_str & dd:mm:ss & Declination \\ 
6 & RAJ2000 & $\arcdeg$ & Right ascension \\ 
7 & err\_RAJ2000 & $\arcdeg$ & Error on RA \\ 
8 & DEJ2000 & $\arcdeg$ & Declination \\ 
9 & err\_DEJ2000 & $\arcdeg$ & Error on Dec \\ 
10 & peak\_flux\_wide & Jy\,beam$^{-1}$ & Peak flux density in wideband image \\ 
11 & err\_peak\_flux\_wide & Jy\,beam$^{-1}$ & Fitting error on peak flux density in wideband image \\ 
12 & int\_flux\_wide & Jy & Integrated flux density in wideband image \\ 
13 & err\_int\_flux\_wide & Jy & Error on integrated flux density in wideband image \\ 
14 & a\_wide & $\arcsec$ & Major axis of source in wideband image \\ 
15 & err\_a\_wide & $\arcsec$ & Error on major axis of source in wideband image \\ 
16 & b\_wide & $\arcsec$ & Minor axis of source in wideband image \\ 
17 & err\_b\_wide & $\arcsec$ & Error on minor axis of source in wideband image \\ 
18 & pa\_wide & $\arcdeg$ & Postion angle of source in wideband image \\ 
19 & err\_pa\_wide & $\arcdeg$ & Error on position angle of source in wideband image \\ 
20 & residual\_mean\_wide & Jy\,beam$^{-1}$ & Mean of residual after source fitting in wideband image \\ 
21 & residual\_std\_wide & Jy\,beam$^{-1}$ & Standard deviation of residual after source fitting \\ 
22 & err\_abs\_flux\_pct & \% & Percent error in absolute flux scale - all frequencies \\ 
23 & err\_fit\_flux\_pct & \% & Percent error on internal flux scale - all frequencies \\
24 & psf\_a\_wide & $\arcsec$ & Major axis of PSF at location of source in wideband image \\
25 & psf\_b\_wide & $\arcsec$ & Minor axis of PSF at location of source in wideband image \\
26 & psf\_pa\_wide & $\arcdeg$ & Position angle of PSF at location of source in wideband image \\
27 & background\_076 & Jy\,beam$^{-1}$ & Background at 76\,MHz \\
28 & local\_rms\_076 & Jy\,beam$^{-1}$ & Local RMS at 76\,MHz \\
29 & peak\_flux\_076 & Jy\,beam$^{-1}$ & Peak flux density at 76\,MHz \\
30 & err\_peak\_flux\_076 & Jy\,beam$^{-1}$ & Fitting error on peak flux density at 76\,MHz \\
31 & int\_flux\_076 & Jy & Integrated flux density at 76\,MHz \\
32 & err\_int\_flux\_076 & Jy & Fitting error on integrated flux density at 76\,MHz \\
33 & a\_076 & $\arcsec$ & Major axis of source at 76\,MHz \\
34 & b\_076 & $\arcsec$ & Minor axis of source at 76\,MHz \\
35 & pa\_076 & $\arcdeg$ & Position angle of source at 76\,MHz \\
36 & residual\_mean\_076 & Jy\,beam$^{-1}$ & Mean of residual after source fitting at 76\,MHz \\
37 & residual\_std\_076 & Jy\,beam$^{-1}$ & Standard deviation of residual after source fitting at 76\,MHz \\
38 & psf\_a\_076 & $\arcsec$ & Major axis of PSF at location of source at 76\,MHz \\
39 & psf\_b\_076 & $\arcsec$ & Minor axis of PSF at location of source at 76\,MHz \\
40 & psf\_pa\_076 & $\arcdeg$ & Position angle of PSF at location of source at 76\,MHz \\
41 & background\_084 & Jy\,beam$^{-1}$ & Background at 84\,MHz \\
42 & local\_rms\_084 & Jy\,beam$^{-1}$ & Local RMS at 84\,MHz \\
43 & peak\_flux\_084 & Jy\,beam$^{-1}$ & Peak flux density at 84\,MHz \\
44 & err\_peak\_flux\_084 & Jy\,beam$^{-1}$ & Fitting error on peak flux density at 84\,MHz \\
45 & int\_flux\_084 & Jy & Integrated flux density at 84\,MHz \\
46 & err\_int\_flux\_084 & Jy & Fitting error on integrated flux density at 84\,MHz \\
47 & a\_084 & $\arcsec$ & Major axis of source at 84\,MHz \\
48 & b\_084 & $\arcsec$ & Minor axis of source at 84\,MHz \\
49 & pa\_084 & $\arcdeg$ & Position angle of source at 84\,MHz \\
50 & residual\_mean\_084 & Jy\,beam$^{-1}$ & Mean of residual after source fitting at 84\,MHz \\
51 & residual\_std\_084 & Jy\,beam$^{-1}$ & Standard deviation of residual after source fitting at 84\,MHz \\
52 & psf\_a\_084 & $\arcsec$ & Major axis of PSF at location of source at 84\,MHz \\
53 & psf\_b\_084 & $\arcsec$ & Minor axis of PSF at location of source at 84\,MHz \\
54 & psf\_pa\_084 & $\arcdeg$ & Position angle of PSF at location of source at 84\,MHz \\
55 & background\_092 & Jy\,beam$^{-1}$ & Background at 92\,MHz \\
56 & local\_rms\_092 & Jy\,beam$^{-1}$ & Local RMS at 92\,MHz \\
57 & peak\_flux\_092 & Jy\,beam$^{-1}$ & Peak flux density at 92\,MHz \\
58 & err\_peak\_flux\_092 & Jy\,beam$^{-1}$ & Fitting error on peak flux density at 92\,MHz \\
59 & int\_flux\_092 & Jy & Integrated flux density at 92\,MHz \\
60 & err\_int\_flux\_092 & Jy & Fitting error on integrated flux density at 92\,MHz \\
61 & a\_092 & $\arcsec$ & Major axis of source at 92\,MHz \\
62 & b\_092 & $\arcsec$ & Minor axis of source at 92\,MHz \\
63 & pa\_092 & $\arcdeg$ & Position angle of source at 92\,MHz \\
64 & residual\_mean\_092 & Jy\,beam$^{-1}$ & Mean of residual after source fitting at 92\,MHz \\
65 & residual\_std\_092 & Jy\,beam$^{-1}$ & Standard deviation of residual after source fitting at 92\,MHz \\
66 & psf\_a\_092 & $\arcsec$ & Major axis of PSF at location of source at 92\,MHz \\
67 & psf\_b\_092 & $\arcsec$ & Minor axis of PSF at location of source at 92\,MHz \\
68 & psf\_pa\_092 & $\arcdeg$ & Position angle of PSF at location of source at 92\,MHz \\
69 & background\_099 & Jy\,beam$^{-1}$ & Background at 99\,MHz \\
70 & local\_rms\_099 & Jy\,beam$^{-1}$ & Local RMS at 99\,MHz \\
71 & peak\_flux\_099 & Jy\,beam$^{-1}$ & Peak flux density at 99\,MHz \\
72 & err\_peak\_flux\_099 & Jy\,beam$^{-1}$ & Fitting error on peak flux density at 99\,MHz \\
73 & int\_flux\_099 & Jy & Integrated flux density at 99\,MHz \\
74 & err\_int\_flux\_099 & Jy & Fitting error on integrated flux density at 99\,MHz \\
75 & a\_099 & $\arcsec$ & Major axis of source at 99\,MHz \\
76 & b\_099 & $\arcsec$ & Minor axis of source at 99\,MHz \\
77 & pa\_099 & $\arcdeg$ & Position angle of source at 99\,MHz \\
78 & residual\_mean\_099 & Jy\,beam$^{-1}$ & Mean of residual after source fitting at 99\,MHz \\
79 & residual\_std\_099 & Jy\,beam$^{-1}$ & Standard deviation of residual after source fitting at 99\,MHz \\
80 & psf\_a\_099 & $\arcsec$ & Major axis of PSF at location of source at 99\,MHz \\
81 & psf\_b\_099 & $\arcsec$ & Minor axis of PSF at location of source at 99\,MHz \\
82 & psf\_pa\_099 & $\arcdeg$ & Position angle of PSF at location of source at 99\,MHz \\
83 & background\_107 & Jy\,beam$^{-1}$ & Background at 107\,MHz \\
84 & local\_rms\_107 & Jy\,beam$^{-1}$ & Local RMS at 107\,MHz \\
85 & peak\_flux\_107 & Jy\,beam$^{-1}$ & Peak flux density at 107\,MHz \\
86 & err\_peak\_flux\_107 & Jy\,beam$^{-1}$ & Fitting error on peak flux density at 107\,MHz \\
87 & int\_flux\_107 & Jy & Integrated flux density at 107\,MHz \\
88 & err\_int\_flux\_107 & Jy & Fitting error on integrated flux density at 107\,MHz \\
89 & a\_107 & $\arcsec$ & Major axis of source at 107\,MHz \\
90 & b\_107 & $\arcsec$ & Minor axis of source at 107\,MHz \\
91 & pa\_107 & $\arcdeg$ & Position angle of source at 107\,MHz \\
92 & residual\_mean\_107 & Jy\,beam$^{-1}$ & Mean of residual after source fitting at 107\,MHz \\
93 & residual\_std\_107 & Jy\,beam$^{-1}$ & Standard deviation of residual after source fitting at 107\,MHz \\
94 & psf\_a\_107 & $\arcsec$ & Major axis of PSF at location of source at 107\,MHz \\
95 & psf\_b\_107 & $\arcsec$ & Minor axis of PSF at location of source at 107\,MHz \\
96 & psf\_pa\_107 & $\arcdeg$ & Position angle of PSF at location of source at 107\,MHz \\
97 & background\_115 & Jy\,beam$^{-1}$ & Background at 115\,MHz \\
98 & local\_rms\_115 & Jy\,beam$^{-1}$ & Local RMS at 115\,MHz \\
99 & peak\_flux\_115 & Jy\,beam$^{-1}$ & Peak flux density at 115\,MHz \\
100 & err\_peak\_flux\_115 & Jy\,beam$^{-1}$ & Fitting error on peak flux density at 115\,MHz \\
101 & int\_flux\_115 & Jy & Integrated flux density at 115\,MHz \\
102 & err\_int\_flux\_115 & Jy & Fitting error on integrated flux density at 115\,MHz \\
103 & a\_115 & $\arcsec$ & Major axis of source at 115\,MHz \\
104 & b\_115 & $\arcsec$ & Minor axis of source at 115\,MHz \\
105 & pa\_115 & $\arcdeg$ & Position angle of source at 115\,MHz \\
106 & residual\_mean\_115 & Jy\,beam$^{-1}$ & Mean of residual after source fitting at 115\,MHz \\
107 & residual\_std\_115 & Jy\,beam$^{-1}$ & Standard deviation of residual after source fitting at 115\,MHz \\
108 & psf\_a\_115 & $\arcsec$ & Major axis of PSF at location of source at 115\,MHz \\
109 & psf\_b\_115 & $\arcsec$ & Minor axis of PSF at location of source at 115\,MHz \\
110 & psf\_pa\_115 & $\arcdeg$ & Position angle of PSF at location of source at 115\,MHz \\
111 & background\_122 & Jy\,beam$^{-1}$ & Background at 122\,MHz \\
112 & local\_rms\_122 & Jy\,beam$^{-1}$ & Local RMS at 122\,MHz \\
113 & peak\_flux\_122 & Jy\,beam$^{-1}$ & Peak flux density at 122\,MHz \\
114 & err\_peak\_flux\_122 & Jy\,beam$^{-1}$ & Fitting error on peak flux density at 122\,MHz \\
115 & int\_flux\_122 & Jy & Integrated flux density at 122\,MHz \\
116 & err\_int\_flux\_122 & Jy & Fitting error on integrated flux density at 122\,MHz \\
117 & a\_122 & $\arcsec$ & Major axis of source at 122\,MHz \\
118 & b\_122 & $\arcsec$ & Minor axis of source at 122\,MHz \\
119 & pa\_122 & $\arcdeg$ & Position angle of source at 122\,MHz \\
120 & residual\_mean\_122 & Jy\,beam$^{-1}$ & Mean of residual after source fitting at 122\,MHz \\
121 & residual\_std\_122 & Jy\,beam$^{-1}$ & Standard deviation of residual after source fitting at 122\,MHz \\
122 & psf\_a\_122 & $\arcsec$ & Major axis of PSF at location of source at 122\,MHz \\
123 & psf\_b\_122 & $\arcsec$ & Minor axis of PSF at location of source at 122\,MHz \\
124 & psf\_pa\_122 & $\arcdeg$ & Position angle of PSF at location of source at 122\,MHz \\
125 & background\_130 & Jy\,beam$^{-1}$ & Background at 130\,MHz \\
126 & local\_rms\_130 & Jy\,beam$^{-1}$ & Local RMS at 130\,MHz \\
127 & peak\_flux\_130 & Jy\,beam$^{-1}$ & Peak flux density at 130\,MHz \\
128 & err\_peak\_flux\_130 & Jy\,beam$^{-1}$ & Fitting error on peak flux density at 130\,MHz \\
129 & int\_flux\_130 & Jy & Integrated flux density at 130\,MHz \\
130 & err\_int\_flux\_130 & Jy & Fitting error on integrated flux density at 130\,MHz \\
131 & a\_130 & $\arcsec$ & Major axis of source at 130\,MHz \\
132 & b\_130 & $\arcsec$ & Minor axis of source at 130\,MHz \\
133 & pa\_130 & $\arcdeg$ & Position angle of source at 130\,MHz \\
134 & residual\_mean\_130 & Jy\,beam$^{-1}$ & Mean of residual after source fitting at 130\,MHz \\
135 & residual\_std\_130 & Jy\,beam$^{-1}$ & Standard deviation of residual after source fitting at 130\,MHz \\
136 & psf\_a\_130 & $\arcsec$ & Major axis of PSF at location of source at 130\,MHz \\
137 & psf\_b\_130 & $\arcsec$ & Minor axis of PSF at location of source at 130\,MHz \\
138 & psf\_pa\_130 & $\arcdeg$ & Position angle of PSF at location of source at 130\,MHz \\
139 & background\_143 & Jy\,beam$^{-1}$ & Background at 143\,MHz \\
140 & local\_rms\_143 & Jy\,beam$^{-1}$ & Local RMS at 143\,MHz \\
141 & peak\_flux\_143 & Jy\,beam$^{-1}$ & Peak flux density at 143\,MHz \\
142 & err\_peak\_flux\_143 & Jy\,beam$^{-1}$ & Fitting error on peak flux density at 143\,MHz \\
143 & int\_flux\_143 & Jy & Integrated flux density at 143\,MHz \\
144 & err\_int\_flux\_143 & Jy & Fitting error on integrated flux density at 143\,MHz \\
145 & a\_143 & $\arcsec$ & Major axis of source at 143\,MHz \\
146 & b\_143 & $\arcsec$ & Minor axis of source at 143\,MHz \\
147 & pa\_143 & $\arcdeg$ & Position angle of source at 143\,MHz \\
148 & residual\_mean\_143 & Jy\,beam$^{-1}$ & Mean of residual after source fitting at 143\,MHz \\
149 & residual\_std\_143 & Jy\,beam$^{-1}$ & Standard deviation of residual after source fitting at 143\,MHz \\
150 & psf\_a\_143 & $\arcsec$ & Major axis of PSF at location of source at 143\,MHz \\
151 & psf\_b\_143 & $\arcsec$ & Minor axis of PSF at location of source at 143\,MHz \\
152 & psf\_pa\_143 & $\arcdeg$ & Position angle of PSF at location of source at 143\,MHz \\
153 & background\_151 & Jy\,beam$^{-1}$ & Background at 151\,MHz \\
154 & local\_rms\_151 & Jy\,beam$^{-1}$ & Local RMS at 151\,MHz \\
155 & peak\_flux\_151 & Jy\,beam$^{-1}$ & Peak flux density at 151\,MHz \\
156 & err\_peak\_flux\_151 & Jy\,beam$^{-1}$ & Fitting error on peak flux density at 151\,MHz \\
157 & int\_flux\_151 & Jy & Integrated flux density at 151\,MHz \\
158 & err\_int\_flux\_151 & Jy & Fitting error on integrated flux density at 151\,MHz \\
159 & a\_151 & $\arcsec$ & Major axis of source at 151\,MHz \\
160 & b\_151 & $\arcsec$ & Minor axis of source at 151\,MHz \\
161 & pa\_151 & $\arcdeg$ & Position angle of source at 151\,MHz \\
162 & residual\_mean\_151 & Jy\,beam$^{-1}$ & Mean of residual after source fitting at 151\,MHz \\
163 & residual\_std\_151 & Jy\,beam$^{-1}$ & Standard deviation of residual after source fitting at 151\,MHz \\
164 & psf\_a\_151 & $\arcsec$ & Major axis of PSF at location of source at 151\,MHz \\
165 & psf\_b\_151 & $\arcsec$ & Minor axis of PSF at location of source at 151\,MHz \\
166 & psf\_pa\_151 & $\arcdeg$ & Position angle of PSF at location of source at 151\,MHz \\
167 & background\_158 & Jy\,beam$^{-1}$ & Background at 158\,MHz \\
168 & local\_rms\_158 & Jy\,beam$^{-1}$ & Local RMS at 158\,MHz \\
169 & peak\_flux\_158 & Jy\,beam$^{-1}$ & Peak flux density at 158\,MHz \\
170 & err\_peak\_flux\_158 & Jy\,beam$^{-1}$ & Fitting error on peak flux density at 158\,MHz \\
171 & int\_flux\_158 & Jy & Integrated flux density at 158\,MHz \\
172 & err\_int\_flux\_158 & Jy & Fitting error on integrated flux density at 158\,MHz \\
173 & a\_158 & $\arcsec$ & Major axis of source at 158\,MHz \\
174 & b\_158 & $\arcsec$ & Minor axis of source at 158\,MHz \\
175 & pa\_158 & $\arcdeg$ & Position angle of source at 158\,MHz \\
176 & residual\_mean\_158 & Jy\,beam$^{-1}$ & Mean of residual after source fitting at 158\,MHz \\
177 & residual\_std\_158 & Jy\,beam$^{-1}$ & Standard deviation of residual after source fitting at 158\,MHz \\
178 & psf\_a\_158 & $\arcsec$ & Major axis of PSF at location of source at 158\,MHz \\
179 & psf\_b\_158 & $\arcsec$ & Minor axis of PSF at location of source at 158\,MHz \\
180 & psf\_pa\_158 & $\arcdeg$ & Position angle of PSF at location of source at 158\,MHz \\
181 & background\_166 & Jy\,beam$^{-1}$ & Background at 166\,MHz \\
182 & local\_rms\_166 & Jy\,beam$^{-1}$ & Local RMS at 166\,MHz \\
183 & peak\_flux\_166 & Jy\,beam$^{-1}$ & Peak flux density at 166\,MHz \\
184 & err\_peak\_flux\_166 & Jy\,beam$^{-1}$ & Fitting error on peak flux density at 166\,MHz \\
185 & int\_flux\_166 & Jy & Integrated flux density at 166\,MHz \\
186 & err\_int\_flux\_166 & Jy & Fitting error on integrated flux density at 166\,MHz \\
187 & a\_166 & $\arcsec$ & Major axis of source at 166\,MHz \\
188 & b\_166 & $\arcsec$ & Minor axis of source at 166\,MHz \\
189 & pa\_166 & $\arcdeg$ & Position angle of source at 166\,MHz \\
190 & residual\_mean\_166 & Jy\,beam$^{-1}$ & Mean of residual after source fitting at 166\,MHz \\
191 & residual\_std\_166 & Jy\,beam$^{-1}$ & Standard deviation of residual after source fitting at 166\,MHz \\
192 & psf\_a\_166 & $\arcsec$ & Major axis of PSF at location of source at 166\,MHz \\
193 & psf\_b\_166 & $\arcsec$ & Minor axis of PSF at location of source at 166\,MHz \\
194 & psf\_pa\_166 & $\arcdeg$ & Position angle of PSF at location of source at 166\,MHz \\
195 & background\_174 & Jy\,beam$^{-1}$ & Background at 174\,MHz \\
196 & local\_rms\_174 & Jy\,beam$^{-1}$ & Local RMS at 174\,MHz \\
197 & peak\_flux\_174 & Jy\,beam$^{-1}$ & Peak flux density at 174\,MHz \\
198 & err\_peak\_flux\_174 & Jy\,beam$^{-1}$ & Fitting error on peak flux density at 174\,MHz \\
199 & int\_flux\_174 & Jy & Integrated flux density at 174\,MHz \\
200 & err\_int\_flux\_174 & Jy & Fitting error on integrated flux density at 174\,MHz \\
201 & a\_174 & $\arcsec$ & Major axis of source at 174\,MHz \\
202 & b\_174 & $\arcsec$ & Minor axis of source at 174\,MHz \\
203 & pa\_174 & $\arcdeg$ & Position angle of source at 174\,MHz \\
204 & residual\_mean\_174 & Jy\,beam$^{-1}$ & Mean of residual after source fitting at 174\,MHz \\
205 & residual\_std\_174 & Jy\,beam$^{-1}$ & Standard deviation of residual after source fitting at 174\,MHz \\
206 & psf\_a\_174 & $\arcsec$ & Major axis of PSF at location of source at 174\,MHz \\
207 & psf\_b\_174 & $\arcsec$ & Minor axis of PSF at location of source at 174\,MHz \\
208 & psf\_pa\_174 & $\arcdeg$ & Position angle of PSF at location of source at 174\,MHz \\
209 & background\_181 & Jy\,beam$^{-1}$ & Background at 181\,MHz \\
210 & local\_rms\_181 & Jy\,beam$^{-1}$ & Local RMS at 181\,MHz \\
211 & peak\_flux\_181 & Jy\,beam$^{-1}$ & Peak flux density at 181\,MHz \\
212 & err\_peak\_flux\_181 & Jy\,beam$^{-1}$ & Fitting error on peak flux density at 181\,MHz \\
213 & int\_flux\_181 & Jy & Integrated flux density at 181\,MHz \\
214 & err\_int\_flux\_181 & Jy & Fitting error on integrated flux density at 181\,MHz \\
215 & a\_181 & $\arcsec$ & Major axis of source at 181\,MHz \\
216 & b\_181 & $\arcsec$ & Minor axis of source at 181\,MHz \\
217 & pa\_181 & $\arcdeg$ & Position angle of source at 181\,MHz \\
218 & residual\_mean\_181 & Jy\,beam$^{-1}$ & Mean of residual after source fitting at 181\,MHz \\
219 & residual\_std\_181 & Jy\,beam$^{-1}$ & Standard deviation of residual after source fitting at 181\,MHz \\
220 & psf\_a\_181 & $\arcsec$ & Major axis of PSF at location of source at 181\,MHz \\
221 & psf\_b\_181 & $\arcsec$ & Minor axis of PSF at location of source at 181\,MHz \\
222 & psf\_pa\_181 & $\arcdeg$ & Position angle of PSF at location of source at 181\,MHz \\
223 & background\_189 & Jy\,beam$^{-1}$ & Background at 189\,MHz \\
224 & local\_rms\_189 & Jy\,beam$^{-1}$ & Local RMS at 189\,MHz \\
225 & peak\_flux\_189 & Jy\,beam$^{-1}$ & Peak flux density at 189\,MHz \\
226 & err\_peak\_flux\_189 & Jy\,beam$^{-1}$ & Fitting error on peak flux density at 189\,MHz \\
227 & int\_flux\_189 & Jy & Integrated flux density at 189\,MHz \\
228 & err\_int\_flux\_189 & Jy & Fitting error on integrated flux density at 189\,MHz \\
229 & a\_189 & $\arcsec$ & Major axis of source at 189\,MHz \\
230 & b\_189 & $\arcsec$ & Minor axis of source at 189\,MHz \\
231 & pa\_189 & $\arcdeg$ & Position angle of source at 189\,MHz \\
232 & residual\_mean\_189 & Jy\,beam$^{-1}$ & Mean of residual after source fitting at 189\,MHz \\
233 & residual\_std\_189 & Jy\,beam$^{-1}$ & Standard deviation of residual after source fitting at 189\,MHz \\
234 & psf\_a\_189 & $\arcsec$ & Major axis of PSF at location of source at 189\,MHz \\
235 & psf\_b\_189 & $\arcsec$ & Minor axis of PSF at location of source at 189\,MHz \\
236 & psf\_pa\_189 & $\arcdeg$ & Position angle of PSF at location of source at 189\,MHz \\
237 & background\_197 & Jy\,beam$^{-1}$ & Background at 197\,MHz \\
238 & local\_rms\_197 & Jy\,beam$^{-1}$ & Local RMS at 197\,MHz \\
239 & peak\_flux\_197 & Jy\,beam$^{-1}$ & Peak flux density at 197\,MHz \\
240 & err\_peak\_flux\_197 & Jy\,beam$^{-1}$ & Fitting error on peak flux density at 197\,MHz \\
241 & int\_flux\_197 & Jy & Integrated flux density at 197\,MHz \\
242 & err\_int\_flux\_197 & Jy & Fitting error on integrated flux density at 197\,MHz \\
243 & a\_197 & $\arcsec$ & Major axis of source at 197\,MHz \\
244 & b\_197 & $\arcsec$ & Minor axis of source at 197\,MHz \\
245 & pa\_197 & $\arcdeg$ & Position angle of source at 197\,MHz \\
246 & residual\_mean\_197 & Jy\,beam$^{-1}$ & Mean of residual after source fitting at 197\,MHz \\
247 & residual\_std\_197 & Jy\,beam$^{-1}$ & Standard deviation of residual after source fitting at 197\,MHz \\
248 & psf\_a\_197 & $\arcsec$ & Major axis of PSF at location of source at 197\,MHz \\
249 & psf\_b\_197 & $\arcsec$ & Minor axis of PSF at location of source at 197\,MHz \\
250 & psf\_pa\_197 & $\arcdeg$ & Position angle of PSF at location of source at 197\,MHz \\
251 & background\_204 & Jy\,beam$^{-1}$ & Background at 204\,MHz \\
252 & local\_rms\_204 & Jy\,beam$^{-1}$ & Local RMS at 204\,MHz \\
253 & peak\_flux\_204 & Jy\,beam$^{-1}$ & Peak flux density at 204\,MHz \\
254 & err\_peak\_flux\_204 & Jy\,beam$^{-1}$ & Fitting error on peak flux density at 204\,MHz \\
255 & int\_flux\_204 & Jy & Integrated flux density at 204\,MHz \\
256 & err\_int\_flux\_204 & Jy & Fitting error on integrated flux density at 204\,MHz \\
257 & a\_204 & $\arcsec$ & Major axis of source at 204\,MHz \\
258 & b\_204 & $\arcsec$ & Minor axis of source at 204\,MHz \\
259 & pa\_204 & $\arcdeg$ & Position angle of source at 204\,MHz \\
260 & residual\_mean\_204 & Jy\,beam$^{-1}$ & Mean of residual after source fitting at 204\,MHz \\
261 & residual\_std\_204 & Jy\,beam$^{-1}$ & Standard deviation of residual after source fitting at 204\,MHz \\
262 & psf\_a\_204 & $\arcsec$ & Major axis of PSF at location of source at 204\,MHz \\
263 & psf\_b\_204 & $\arcsec$ & Minor axis of PSF at location of source at 204\,MHz \\
264 & psf\_pa\_204 & $\arcdeg$ & Position angle of PSF at location of source at 204\,MHz \\
265 & background\_212 & Jy\,beam$^{-1}$ & Background at 212\,MHz \\
266 & local\_rms\_212 & Jy\,beam$^{-1}$ & Local RMS at 212\,MHz \\
267 & peak\_flux\_212 & Jy\,beam$^{-1}$ & Peak flux density at 212\,MHz \\
268 & err\_peak\_flux\_212 & Jy\,beam$^{-1}$ & Fitting error on peak flux density at 212\,MHz \\
269 & int\_flux\_212 & Jy & Integrated flux density at 212\,MHz \\
270 & err\_int\_flux\_212 & Jy & Fitting error on integrated flux density at 212\,MHz \\
271 & a\_212 & $\arcsec$ & Major axis of source at 212\,MHz \\
272 & b\_212 & $\arcsec$ & Minor axis of source at 212\,MHz \\
273 & pa\_212 & $\arcdeg$ & Position angle of source at 212\,MHz \\
274 & residual\_mean\_212 & Jy\,beam$^{-1}$ & Mean of residual after source fitting at 212\,MHz \\
275 & residual\_std\_212 & Jy\,beam$^{-1}$ & Standard deviation of residual after source fitting at 212\,MHz \\
276 & psf\_a\_212 & $\arcsec$ & Major axis of PSF at location of source at 212\,MHz \\
277 & psf\_b\_212 & $\arcsec$ & Minor axis of PSF at location of source at 212\,MHz \\
278 & psf\_pa\_212 & $\arcdeg$ & Position angle of PSF at location of source at 212\,MHz \\
279 & background\_220 & Jy\,beam$^{-1}$ & Background at 220\,MHz \\
280 & local\_rms\_220 & Jy\,beam$^{-1}$ & Local RMS at 220\,MHz \\
281 & peak\_flux\_220 & Jy\,beam$^{-1}$ & Peak flux density at 220\,MHz \\
282 & err\_peak\_flux\_220 & Jy\,beam$^{-1}$ & Fitting error on peak flux density at 220\,MHz \\
283 & int\_flux\_220 & Jy & Integrated flux density at 220\,MHz \\
284 & err\_int\_flux\_220 & Jy & Fitting error on integrated flux density at 220\,MHz \\
285 & a\_220 & $\arcsec$ & Major axis of source at 220\,MHz \\
286 & b\_220 & $\arcsec$ & Minor axis of source at 220\,MHz \\
287 & pa\_220 & $\arcdeg$ & Position angle of source at 220\,MHz \\
288 & residual\_mean\_220 & Jy\,beam$^{-1}$ & Mean of residual after source fitting at 220\,MHz \\
289 & residual\_std\_220 & Jy\,beam$^{-1}$ & Standard deviation of residual after source fitting at 220\,MHz \\
290 & psf\_a\_220 & $\arcsec$ & Major axis of PSF at location of source at 220\,MHz \\
291 & psf\_b\_220 & $\arcsec$ & Minor axis of PSF at location of source at 220\,MHz \\
292 & psf\_pa\_220 & $\arcdeg$ & Position angle of PSF at location of source at 220\,MHz \\
293 & background\_227 & Jy\,beam$^{-1}$ & Background at 227\,MHz \\
294 & local\_rms\_227 & Jy\,beam$^{-1}$ & Local RMS at 227\,MHz \\
295 & peak\_flux\_227 & Jy\,beam$^{-1}$ & Peak flux density at 227\,MHz \\
296 & err\_peak\_flux\_227 & Jy\,beam$^{-1}$ & Fitting error on peak flux density at 227\,MHz \\
297 & int\_flux\_227 & Jy & Integrated flux density at 227\,MHz \\
298 & err\_int\_flux\_227 & Jy & Fitting error on integrated flux density at 227\,MHz \\
299 & a\_227 & $\arcsec$ & Major axis of source at 227\,MHz \\
300 & b\_227 & $\arcsec$ & Minor axis of source at 227\,MHz \\
301 & pa\_227 & $\arcdeg$ & Position angle of source at 227\,MHz \\
302 & residual\_mean\_227 & Jy\,beam$^{-1}$ & Mean of residual after source fitting at 227\,MHz \\
303 & residual\_std\_227 & Jy\,beam$^{-1}$ & Standard deviation of residual after source fitting at 227\,MHz \\
304 & psf\_a\_227 & $\arcsec$ & Major axis of PSF at location of source at 227\,MHz \\
305 & psf\_b\_227 & $\arcsec$ & Minor axis of PSF at location of source at 227\,MHz \\
306 & psf\_pa\_227 & $\arcdeg$ & Position angle of PSF at location of source at 227\,MHz \\
307 & background\_W\_087 & Jy\,beam$^{-1}$ & Background at 072-103\,MHz \\
308 & local\_rms\_W\_087 & Jy\,beam$^{-1}$ & Local RMS at 072-103\,MHz \\
309 & peak\_flux\_W\_087 & Jy\,beam$^{-1}$ & Peak flux density at 072-103\,MHz \\
310 & err\_peak\_flux\_W\_087 & Jy\,beam$^{-1}$ & Fitting error on peak flux density at 072-103\,MHz \\
311 & int\_flux\_W\_087 & Jy & Integrated flux density at 072-103\,MHz \\
312 & err\_int\_flux\_W\_087 & Jy & Fitting error on integrated flux density at 072-103\,MHz \\
313 & a\_W\_087 & $\arcsec$ & Major axis of source at 072-103\,MHz \\
314 & b\_W\_087 & $\arcsec$ & Minor axis of source at 072-103\,MHz \\
315 & pa\_W\_087 & $\arcdeg$ & Position angle of source at 072-103\,MHz \\
316 & residual\_mean\_W\_087 & Jy\,beam$^{-1}$ & Mean of residual after source fitting at 072-103\,MHz \\
317 & residual\_std\_W\_087 & Jy\,beam$^{-1}$ & Standard deviation of residual after source fitting at 072-103\,MHz \\
318 & psf\_a\_W\_087 & $\arcsec$ & Major axis of PSF at location of source at 072-103\,MHz \\
319 & psf\_b\_W\_087 & $\arcsec$ & Minor axis of PSF at location of source at 072-103\,MHz \\
320 & psf\_pa\_W\_087 & $\arcdeg$ & Position angle of PSF at location of source at 072-103\,MHz \\
321 & background\_W\_118 & Jy\,beam$^{-1}$ & Background at 103-134\,MHz \\
322 & local\_rms\_W\_118 & Jy\,beam$^{-1}$ & Local RMS at 103-134\,MHz \\
323 & peak\_flux\_W\_118 & Jy\,beam$^{-1}$ & Peak flux density at 103-134\,MHz \\
324 & err\_peak\_flux\_W\_118 & Jy\,beam$^{-1}$ & Fitting error on peak flux density at 103-134\,MHz \\
325 & int\_flux\_W\_118 & Jy & Integrated flux density at 103-134\,MHz \\
326 & err\_int\_flux\_W\_118 & Jy & Fitting error on integrated flux density at 103-134\,MHz \\
327 & a\_W\_118 & $\arcsec$ & Major axis of source at 103-134\,MHz \\
328 & b\_W\_118 & $\arcsec$ & Minor axis of source at 103-134\,MHz \\
329 & pa\_W\_118 & $\arcdeg$ & Position angle of source at 103-134\,MHz \\
330 & residual\_mean\_W\_118 & Jy\,beam$^{-1}$ & Mean of residual after source fitting at 103-134\,MHz \\
331 & residual\_std\_W\_118 & Jy\,beam$^{-1}$ & Standard deviation of residual after source fitting at 103-134\,MHz \\
332 & psf\_a\_W\_118 & $\arcsec$ & Major axis of PSF at location of source at 103-134\,MHz \\
333 & psf\_b\_W\_118 & $\arcsec$ & Minor axis of PSF at location of source at 103-134\,MHz \\
334 & psf\_pa\_W\_118 & $\arcsec$ & Postion angle of PSF at location of source at 103-134\,MHz \\
335 & background\_W\_154 & Jy\,beam$^{-1}$ & Background at 139-170\,MHz \\
336 & local\_rms\_W\_154 & Jy\,beam$^{-1}$ & Local RMS at 139-170\,MHz \\
337 & peak\_flux\_W\_154 & Jy\,beam$^{-1}$ & Peak flux density at 139-170\,MHz \\
338 & err\_peak\_flux\_W\_154 & Jy\,beam$^{-1}$ & Fitting error on peak flux density at 139-170\,MHz \\
339 & int\_flux\_W\_154 & Jy & Integrated flux density at 139-170\,MHz \\
340 & err\_int\_flux\_W\_154 & Jy & Fitting error on integrated flux density at 139-170\,MHz \\
341 & a\_W\_154 & $\arcsec$ & Major axis of source at 139-170\,MHz \\
342 & b\_W\_154 & $\arcsec$ & Minor axis of source at 139-170\,MHz \\
343 & pa\_W\_154 & $\arcdeg$ & Position angle of source at 139-170\,MHz \\
344 & residual\_mean\_W\_154 & Jy\,beam$^{-1}$ & Mean of residual after source fitting at 139-170\,MHz \\
345 & residual\_std\_W\_154 & Jy\,beam$^{-1}$ & Standard deviation of residual after source fitting at 139-170\,MHz \\
346 & psf\_a\_W\_154 & $\arcsec$ & Major axis of PSF at location of source at 139-170\,MHz \\
347 & psf\_b\_W\_154 & $\arcsec$ & Minor axis of PSF at location of source at 139-170\,MHz \\
348 & psf\_pa\_W\_154 & $\arcsec$ & Postion angle of PSF at location of source at 139-170\,MHz \\
349 & background\_W\_185 & Jy\,beam$^{-1}$ & Background at 170-200\,MHz \\
350 & local\_rms\_W\_185 & Jy\,beam$^{-1}$ & Local RMS at 170-200\,MHz \\
351 & peak\_flux\_W\_185 & Jy\,beam$^{-1}$ & Peak flux density at 170-200\,MHz \\
352 & err\_peak\_flux\_W\_185 & Jy\,beam$^{-1}$ & Fitting error on peak flux density at 170-200\,MHz \\
353 & int\_flux\_W\_185 & Jy & Integrated flux density at 170-200\,MHz \\
354 & err\_int\_flux\_W\_185 & Jy & Fitting error on integrated flux density at 170-200\,MHz \\
355 & a\_W\_185 & $\arcsec$ & Major axis of source at 170-200\,MHz \\
356 & b\_W\_185 & $\arcsec$ & Minor axis of source at 170-200\,MHz \\
357 & pa\_W\_185 & $\arcdeg$ & Position angle of source at 170-200\,MHz \\
358 & residual\_mean\_W\_185 & Jy\,beam$^{-1}$ & Mean of residual after source fitting at 170-200\,MHz \\
359 & residual\_std\_W\_185 & Jy\,beam$^{-1}$ & Standard deviation of residual after source fitting at 170-200\,MHz \\
360 & psf\_a\_W\_185 & $\arcsec$ & Major axis of PSF at location of source at 170-200\,MHz \\
361 & psf\_b\_W\_185 & $\arcsec$ & Minor axis of PSF at location of source at 170-200\,MHz \\
362 & psf\_pa\_W\_185 & $\arcsec$ & Postion angle of PSF at location of source at 170-200\,MHz \\
363 & background\_W\_215 & Jy\,beam$^{-1}$ & Background at 200-231\,MHz \\
364 & local\_rms\_W\_215 & Jy\,beam$^{-1}$ & Local RMS at 200-231\,MHz \\
365 & peak\_flux\_W\_215 & Jy\,beam$^{-1}$ & Peak flux density at 200-231\,MHz \\
366 & err\_peak\_flux\_W\_215 & Jy\,beam$^{-1}$ & Fitting error on peak flux density at 200-231\,MHz \\
367 & int\_flux\_W\_215 & Jy & Integrated flux density at 200-231\,MHz \\
368 & err\_int\_flux\_W\_215 & Jy & Fitting error on integrated flux density at 200-231\,MHz \\
369 & a\_W\_215 & $\arcsec$ & Major axis of source at 200-231\,MHz \\
370 & b\_W\_215 & $\arcsec$ & Minor axis of source at 200-231\,MHz \\
371 & pa\_W\_215 & $\arcdeg$ & Position angle of source at 200-231\,MHz \\
372 & residual\_mean\_W\_215 & Jy\,beam$^{-1}$ & Mean of residual after source fitting at 200-231\,MHz \\
373 & residual\_std\_W\_215 & Jy\,beam$^{-1}$ & Standard deviation of residual after source fitting at 200-231\,MHz \\
374 & psf\_a\_W\_215 & $\arcsec$ & Major axis of PSF at location of source at 200-231\,MHz \\
375 & psf\_b\_W\_215 & $\arcsec$ & Minor axis of PSF at location of source at 200-231\,MHz \\
376 & psf\_pa\_W\_215 & $\arcsec$ & Postion angle of PSF at location of source at 200-231\,MHz \\
377 & sp\_int\_flux\_fit\_200 & Jy & Power-law fitted flux density at 200\,MHz \\
378 & err\_sp\_int\_flux\_fit\_200 & Jy & Error on power-law fitted flux density at 200\,MHz \\
379 & sp\_alpha & -- & Fitted spectral index assuming a power-law SED \\
380 & err\_sp\_alpha & -- & Error on power-law fitted spectral index \\
381 & sp\_reduced\_chi2 & -- & Reduced $\chi^2$ statistic for power-law SED fit \\
382 & csp\_int\_flux\_fit\_200 & Jy & Curved SED fitted flux density at 200\,MHz \\
383 & err\_csp\_int\_flux\_fit\_200 & Jy & Error on curved SED fitted flux density at 200\,MHz \\
384 & csp\_alpha & -- & Fitted spectral index assuming a curved SED \\
385 & err\_csp\_alpha & -- & Error on curved SED fitted spectral index \\
386 & csp\_beta & -- & Fitted curvature index for curved SED fit \\
387 & err\_csp\_beta & -- & Error on curvature index for curved SED fit \\
388 & csp\_reduced\_chi2 & -- & Reduced $\chi^2$ statistic for curved SED fit \\
\end{supertabular}


\end{appendix}

\twocolumn

\bibliographystyle{pasa-mnras}
\bibliography{refs}

\begin{thebibliography}{}
\makeatletter
\relax
\def\mn@urlcharsother{\let\do\@makeother \do\$\do\&\do\#\do\^\do\_\do\%\do\~}
\definecolor{darkblue}{rgb}{0,0,0.597656}
\def\mndoi{\begingroup\mn@urlcharsother \@ifnextchar [ {\mndoi@} {\mndoi@[]}}
\def\mndoi@[#1]#2{\def\@tempa{#1}\ifx\@tempa\@empty \href
  {http://dx.doi.org/#2} {\textcolor{darkblue}{doi:#2}}\else \href
  {http://dx.doi.org/#2} {\textcolor{darkblue}{#1}}\fi \endgroup}
\def\mn@eprint#1#2{\mn@eprint@#1:#2::\@nil}
\def\mn@eprint@arXiv#1{\href {http://arxiv.org/abs/#1} {{\tt arXiv:#1}}}
\def\mn@eprint@dblp#1{\href {http://dblp.uni-trier.de/rec/bibtex/#1.xml}
  {dblp:#1}}
\def\mn@eprint@#1:#2:#3:#4\@nil{\def\@tempa {#1}\def\@tempb {#2}\def\@tempc
  {#3}\ifx \@tempc \@empty \let \@tempc \@tempb \let \@tempb \@tempa \fi \ifx
  \@tempb \@empty \def\@tempb {arXiv}\fi \@ifundefined
  {mn@eprint@\@tempb}{\@tempb:\@tempc}{\expandafter \expandafter \csname
  mn@eprint@\@tempb\endcsname \expandafter{\@tempc}}}

\bibitem[\protect\citeauthoryear{{Adams} et~al.,}{{Adams}
  et~al.}{2022}]{2022A&A...667A..38A}
{Adams} E.~A.~K.,  et~al., 2022, \mndoi [\aap] {10.1051/0004-6361/202244007},
  \href {https://ui.adsabs.harvard.edu/abs/2022A&A...667A..38A} {667, A38}

\bibitem[\protect\citeauthoryear{{Astropy Collaboration}, {Robitaille},
  {Tollerud}, {Greenfield}, {Droettboom}, {Bray}  et~al.}{{Astropy
  Collaboration} et~al.}{2013}]{Astropy}
{Astropy Collaboration} {Robitaille} T.~P.,  {Tollerud} E.~J.,  {Greenfield}
  P.,  {Droettboom} M.,  {Bray} E.,   et~al., 2013, \mndoi [\aap]
  {10.1051/0004-6361/201322068}, \href
  {https://ui.adsabs.harvard.edu/abs/2013A&A...558A..33A} {558, A33}

\bibitem[\protect\citeauthoryear{{Beardsley} et~al.,}{{Beardsley}
  et~al.}{2019}]{2019PASA...36...50B}
{Beardsley} A.~P.,  et~al., 2019, \mndoi [\pasa] {10.1017/pasa.2019.41}, \href
  {https://ui.adsabs.harvard.edu/abs/2019PASA...36...50B} {36, e050}

\bibitem[\protect\citeauthoryear{{Bertin}, {Mellier}, {Radovich}, {Missonnier},
  {Didelon}  \& {Morin}}{{Bertin} et~al.}{2002}]{2002ASPC..281..228B}
{Bertin} E.,  {Mellier} Y.,  {Radovich} M.,  {Missonnier} G.,  {Didelon} P.,
  {Morin} B.,  2002, in {Bohlender} D.~A.,  {Durand} D.,   {Handley} T.~H.,
  eds,  Astronomical Society of the Pacific Conference Series Vol. 281,
  Astronomical Data Analysis Software and Systems XI. p.~228

\bibitem[\protect\citeauthoryear{{Best} et~al.,}{{Best}
  et~al.}{2023}]{2023MNRAS.523.1729B}
{Best} P.~N.,  et~al., 2023, \mndoi [\mnras] {10.1093/mnras/stad1308}, \href
  {https://ui.adsabs.harvard.edu/abs/2023MNRAS.523.1729B} {523, 1729}

\bibitem[\protect\citeauthoryear{{Briggs}}{{Briggs}}{1995}]{1995AAS...18711202B}
{Briggs} D.~S.,  1995, in American Astronomical Society Meeting Abstracts. p.
  112.02

\bibitem[\protect\citeauthoryear{{Callingham} et~al.,}{{Callingham}
  et~al.}{2017}]{2017ApJ...836..174C}
{Callingham} J.~R.,  et~al., 2017, \mndoi [\apj] {10.3847/1538-4357/836/2/174},
  \href {https://ui.adsabs.harvard.edu/abs/2017ApJ...836..174C} {836, 174}

\bibitem[\protect\citeauthoryear{{Condon}, {Cotton}, {Greisen}, {Yin},
  {Perley}, {Taylor}  \& {Broderick}}{{Condon}
  et~al.}{1998}]{1998AJ....115.1693C}
{Condon} J.~J.,  {Cotton} W.~D.,  {Greisen} E.~W.,  {Yin} Q.~F.,  {Perley}
  R.~A.,  {Taylor} G.~B.,   {Broderick} J.~J.,  1998, \mndoi [\aj]
  {10.1086/300337}, \href {http://adsabs.harvard.edu/abs/1998AJ....115.1693C}
  {115, 1693}

\bibitem[\protect\citeauthoryear{{Deka} et~al.,}{{Deka}
  et~al.}{2023}]{2023arXiv230812347D}
{Deka} P.~P.,  et~al., 2023, \mndoi [arXiv e-prints]
  {10.48550/arXiv.2308.12347}, \href
  {https://ui.adsabs.harvard.edu/abs/2023arXiv230812347D} {p. arXiv:2308.12347}

\bibitem[\protect\citeauthoryear{{Dubois}, {Hinsen}  \& {Hugunin}}{{Dubois}
  et~al.}{1996}]{NumPy}
{Dubois} P.~F.,  {Hinsen} K.,   {Hugunin} J.,  1996, Comput. Phys. Commun.,
  \href {https://ui.adsabs.harvard.edu/abs/1996ComPh..10..262D} {10, 262}

\bibitem[\protect\citeauthoryear{{Duchesne}, {Johnston-Hollitt}, {Zhu}, {Wayth}
   \& {Line}}{{Duchesne} et~al.}{2020}]{2020PASA...37...37D}
{Duchesne} S.~W.,  {Johnston-Hollitt} M.,  {Zhu} Z.,  {Wayth} R.~B.,   {Line}
  J.~L.~B.,  2020, \mndoi [\pasa] {10.1017/pasa.2020.29}, \href
  {https://ui.adsabs.harvard.edu/abs/2020PASA...37...37D} {37, e037}

\bibitem[\protect\citeauthoryear{{Duchesne} et~al.,}{{Duchesne}
  et~al.}{2023}]{2023PASA...40...34D}
{Duchesne} S.~W.,  et~al., 2023, \mndoi [\pasa] {10.1017/pasa.2023.31}, \href
  {https://ui.adsabs.harvard.edu/abs/2023PASA...40...34D} {40, e034}

\bibitem[\protect\citeauthoryear{{Duffy} \& {Blundell}}{{Duffy} \&
  {Blundell}}{2012}]{2012MNRAS.421..108D}
{Duffy} P.,  {Blundell} K.~M.,  2012, \mndoi [\mnras]
  {10.1111/j.1365-2966.2011.20239.x}, \href
  {https://ui.adsabs.harvard.edu/abs/2012MNRAS.421..108D} {421, 108}

\bibitem[\protect\citeauthoryear{{Franzen}, {Vernstrom}, {Jackson},
  {Hurley-Walker}, {Ekers}, {Heald}, {Seymour}  \& {White}}{{Franzen}
  et~al.}{2019}]{2019PASA...36....4F}
{Franzen} T.~M.~O.,  {Vernstrom} T.,  {Jackson} C.~A.,  {Hurley-Walker} N.,
  {Ekers} R.~D.,  {Heald} G.,  {Seymour} N.,   {White} S.~V.,  2019, \mndoi
  [\pasa] {10.1017/pasa.2018.52}, \href
  {https://ui.adsabs.harvard.edu/abs/2019PASA...36....4F} {36, e004}

\bibitem[\protect\citeauthoryear{{Franzen}, {Hurley-Walker}, {White},
  {Hancock}, {Seymour}, {Kapi{\'n}ska}, {Staveley-Smith}  \& {Wayth}}{{Franzen}
  et~al.}{2021}]{2021PASA...38...14F}
{Franzen} T.~M.~O.,  {Hurley-Walker} N.,  {White} S.~V.,  {Hancock} P.~J.,
  {Seymour} N.,  {Kapi{\'n}ska} A.~D.,  {Staveley-Smith} L.,   {Wayth} R.~B.,
  2021, \mndoi [\pasa] {10.1017/pasa.2021.5}, \href
  {https://ui.adsabs.harvard.edu/abs/2021PASA...38...14F} {38, e014}

\bibitem[\protect\citeauthoryear{{Hale} et~al.,}{{Hale}
  et~al.}{2024}]{2024MNRAS.527.6540H}
{Hale} C.~L.,  et~al., 2024, \mndoi [\mnras] {10.1093/mnras/stad3088}, \href
  {https://ui.adsabs.harvard.edu/abs/2024MNRAS.527.6540H} {527, 6540}

\bibitem[\protect\citeauthoryear{{Hancock}, {Murphy}, {Gaensler}, {Hopkins}  \&
  {Curran}}{{Hancock} et~al.}{2012}]{2012MNRAS.422.1812H}
{Hancock} P.~J.,  {Murphy} T.,  {Gaensler} B.~M.,  {Hopkins} A.,   {Curran}
  J.~R.,  2012, \mndoi [\mnras] {10.1111/j.1365-2966.2012.20768.x}, \href
  {http://adsabs.harvard.edu/abs/2012MNRAS.422.1812H} {422, 1812}

\bibitem[\protect\citeauthoryear{{Hancock}, {Trott}  \&
  {Hurley-Walker}}{{Hancock} et~al.}{2018}]{2018PASA...35...11H}
{Hancock} P.~J.,  {Trott} C.~M.,   {Hurley-Walker} N.,  2018, \mndoi [\pasa]
  {10.1017/pasa.2018.3}, \href
  {http://adsabs.harvard.edu/abs/2018PASA...35...11H} {35, e011}

\bibitem[\protect\citeauthoryear{Harris et~al.,}{Harris
  et~al.}{2020}]{harris2020array}
Harris C.~R.,  et~al., 2020, Nature, 585, 357

\bibitem[\protect\citeauthoryear{{Heywood} et~al.,}{{Heywood}
  et~al.}{2022}]{2022MNRAS.509.2150H}
{Heywood} I.,  et~al., 2022, \mndoi [\mnras] {10.1093/mnras/stab3021}, \href
  {https://ui.adsabs.harvard.edu/abs/2022MNRAS.509.2150H} {509, 2150}

\bibitem[\protect\citeauthoryear{{Hunter}}{{Hunter}}{2007}]{Matplotlib}
{Hunter} J.~D.,  2007, \mndoi [Comput. Sci. Eng.] {10.1109/MCSE.2007.55}, \href
  {https://ui.adsabs.harvard.edu/abs/2007CSE.....9...90H} {9, 90}

\bibitem[\protect\citeauthoryear{{Hurley-Walker} et~al.,}{{Hurley-Walker}
  et~al.}{2017}]{2017MNRAS.464.1146H}
{Hurley-Walker} N.,  et~al., 2017, \mndoi [\mnras] {10.1093/mnras/stw2337},
  \href {http://adsabs.harvard.edu/abs/2017MNRAS.464.1146H} {464, 1146}

\bibitem[\protect\citeauthoryear{{Hurley-Walker} et~al.,}{{Hurley-Walker}
  et~al.}{2022a}]{2022PASA...39...35H}
{Hurley-Walker} N.,  et~al., 2022a, \mndoi [\pasa] {10.1017/pasa.2022.17},
  \href {https://ui.adsabs.harvard.edu/abs/2022PASA...39...35H} {39, e035}

\bibitem[\protect\citeauthoryear{{Hurley-Walker} et~al.,}{{Hurley-Walker}
  et~al.}{2022b}]{2022Natur.601..526H}
{Hurley-Walker} N.,  et~al., 2022b, \mndoi [\nat]
  {doi.org/10.1038/s41586-021-04272-x}, 601, 526

\bibitem[\protect\citeauthoryear{{Kurtzer}, {Sochat}  \& {Bauer}}{{Kurtzer}
  et~al.}{2017}]{2017PLoSO..1277459K}
{Kurtzer} G.~M.,  {Sochat} V.,   {Bauer} M.~W.,  2017, \mndoi [PLoS ONE]
  {10.1371/journal.pone.0177459}, \href
  {https://ui.adsabs.harvard.edu/abs/2017PLoSO..1277459K} {12, e0177459}

\bibitem[\protect\citeauthoryear{{Lacy} et~al.,}{{Lacy}
  et~al.}{2020}]{2020PASP..132c5001L}
{Lacy} M.,  et~al., 2020, \mndoi [\pasp] {10.1088/1538-3873/ab63eb}, \href
  {https://ui.adsabs.harvard.edu/abs/2020PASP..132c5001L} {132, 035001}

\bibitem[\protect\citeauthoryear{{Mauch}, {Murphy}, {Buttery}, {Curran},
  {Hunstead}, {Piestrzynski}, {Robertson}  \& {Sadler}}{{Mauch}
  et~al.}{2003}]{2003MNRAS.342.1117M}
{Mauch} T.,  {Murphy} T.,  {Buttery} H.~J.,  {Curran} J.,  {Hunstead} R.~W.,
  {Piestrzynski} B.,  {Robertson} J.~G.,   {Sadler} E.~M.,  2003, \mndoi
  [\mnras] {10.1046/j.1365-8711.2003.06605.x}, \href
  {http://adsabs.harvard.edu/abs/2003MNRAS.342.1117M} {342, 1117}

\bibitem[\protect\citeauthoryear{{McConnell} et~al.,}{{McConnell}
  et~al.}{2020}]{2020PASA...37...48M}
{McConnell} D.,  et~al., 2020, \mndoi [\pasa] {10.1017/pasa.2020.41}, \href
  {https://ui.adsabs.harvard.edu/abs/2020PASA...37...48M} {37, e048}

\bibitem[\protect\citeauthoryear{{Morabito} et~al.,}{{Morabito}
  et~al.}{2022}]{2022A&A...658A...1M}
{Morabito} L.~K.,  et~al., 2022, \mndoi [\aap] {10.1051/0004-6361/202140649},
  \href {https://ui.adsabs.harvard.edu/abs/2022A&A...658A...1M} {658, A1}

\bibitem[\protect\citeauthoryear{{Morgan}, {Chhetri}  \& {Ekers}}{{Morgan}
  et~al.}{2022}]{2022PASA...39...63M}
{Morgan} J.~S.,  {Chhetri} R.,   {Ekers} R.,  2022, \mndoi [\pasa]
  {10.1017/pasa.2022.56}, \href
  {https://ui.adsabs.harvard.edu/abs/2022PASA...39...63M} {39, e063}

\bibitem[\protect\citeauthoryear{{Morgan}, {McCauley}, {Waszewski}, {Ekers}  \&
  {Chhetri}}{{Morgan} et~al.}{2023}]{2023SpWea..2103396M}
{Morgan} J.,  {McCauley} P.~I.,  {Waszewski} A.,  {Ekers} R.,   {Chhetri} R.,
  2023, \mndoi [Space Weather] {10.1029/2022SW003396}, \href
  {https://ui.adsabs.harvard.edu/abs/2023SpWea..2103396M} {21, e2022SW003396}

\bibitem[\protect\citeauthoryear{{Offringa} \& {Smirnov}}{{Offringa} \&
  {Smirnov}}{2017}]{2017MNRAS.471..301O}
{Offringa} A.~R.,  {Smirnov} O.,  2017, \mndoi [\mnras]
  {10.1093/mnras/stx1547}, \href
  {https://ui.adsabs.harvard.edu/abs/2017MNRAS.471..301O} {471, 301}

\bibitem[\protect\citeauthoryear{{Offringa}, {van de Gronde}  \&
  {Roerdink}}{{Offringa} et~al.}{2012}]{2012A+A...539A..95O}
{Offringa} A.~R.,  {van de Gronde} J.~J.,   {Roerdink} J.~B.~T.~M.,  2012,
  \mndoi [\aap] {10.1051/0004-6361/201118497}, \href
  {http://cdsads.u-strasbg.fr/abs/2012A%26A...539A..95O} {539, A95}

\bibitem[\protect\citeauthoryear{{Offringa} et~al.,}{{Offringa}
  et~al.}{2014}]{2014MNRAS.444..606O}
{Offringa} A.~R.,  et~al., 2014, \mndoi [\mnras] {10.1093/mnras/stu1368}, \href
  {http://adsabs.harvard.edu/abs/2014MNRAS.444..606O} {444, 606}

\bibitem[\protect\citeauthoryear{{Offringa} et~al.,}{{Offringa}
  et~al.}{2016}]{2016MNRAS.458.1057O}
{Offringa} A.~R.,  et~al., 2016, \mndoi [\mnras] {10.1093/mnras/stw310}, \href
  {http://adsabs.harvard.edu/abs/2016MNRAS.458.1057O} {458, 1057}

\bibitem[\protect\citeauthoryear{{Oliphant}}{{Oliphant}}{2007}]{SciPy}
{Oliphant} T.~E.,  2007, \mndoi [Comput. Sci. Eng.] {10.1109/MCSE.2007.58},
  \href {https://ui.adsabs.harvard.edu/abs/2007CSE.....9c..10O} {9, 10}

\bibitem[\protect\citeauthoryear{{Riseley} et~al.,}{{Riseley}
  et~al.}{2018}]{2018PASA...35...43R}
{Riseley} C.~J.,  et~al., 2018, \mndoi [\pasa] {10.1017/pasa.2018.39}, \href
  {https://ui.adsabs.harvard.edu/abs/2018PASA...35...43R} {35, e043}

\bibitem[\protect\citeauthoryear{{Riseley} et~al.,}{{Riseley}
  et~al.}{2020}]{2020PASA...37...29R}
{Riseley} C.~J.,  et~al., 2020, \mndoi [\pasa] {10.1017/pasa.2020.20}, \href
  {https://ui.adsabs.harvard.edu/abs/2020PASA...37...29R} {37, e029}

\bibitem[\protect\citeauthoryear{{Ross}, {Hurley-Walker}, {Seymour},
  {Callingham}, {Galvin}  \& {Johnston-Hollitt}}{{Ross}
  et~al.}{2022}]{2022MNRAS.512.5358R}
{Ross} K.,  {Hurley-Walker} N.,  {Seymour} N.,  {Callingham} J.~R.,  {Galvin}
  T.~J.,   {Johnston-Hollitt} M.,  2022, \mndoi [\mnras]
  {10.1093/mnras/stac819}, \href
  {https://ui.adsabs.harvard.edu/abs/2022MNRAS.512.5358R} {512, 5358}

\bibitem[\protect\citeauthoryear{{Sault}, {Teuben}  \& {Wright}}{{Sault}
  et~al.}{1995}]{Miriad}
{Sault} R.~J.,  {Teuben} P.~J.,   {Wright} M.~C.~H.,  1995, in {Shaw} R.~A.,
  {Payne} H.~E.,   {Hayes} J.~J.~E.,  eds,  ASP Conference Series Vol. 77,
  Astronomical Data Analysis Software and Systems IV. p.~433 (\mn@eprint
  {arXiv} {astro-ph/0612759})

\bibitem[\protect\citeauthoryear{{Shimwell} et~al.,}{{Shimwell}
  et~al.}{2017}]{2017A&A...598A.104S}
{Shimwell} T.~W.,  et~al., 2017, \mndoi [\aap] {10.1051/0004-6361/201629313},
  \href {https://ui.adsabs.harvard.edu/abs/2017A&A...598A.104S} {598, A104}

\bibitem[\protect\citeauthoryear{{Shimwell} et~al.,}{{Shimwell}
  et~al.}{2022}]{2022A&A...659A...1S}
{Shimwell} T.~W.,  et~al., 2022, \mndoi [\aap] {10.1051/0004-6361/202142484},
  \href {https://ui.adsabs.harvard.edu/abs/2022A&A...659A...1S} {659, A1}

\bibitem[\protect\citeauthoryear{{Taylor}}{{Taylor}}{2005}]{Topcat}
{Taylor} M.~B.,  2005, in {Shopbell} P.,  {Britton} M.,   {Ebert} R.,  eds,
  ASP Conference Series Vol. 347, Astronomical Data Analysis Software and
  Systems XIV. p.~29

\bibitem[\protect\citeauthoryear{{Thomson} et~al.,}{{Thomson}
  et~al.}{2023}]{2023PASA...40...40T}
{Thomson} A. J.~M.,  et~al., 2023, \mndoi [\pasa] {10.1017/pasa.2023.38}, \href
  {https://ui.adsabs.harvard.edu/abs/2023PASA...40...40T} {40, e040}

\bibitem[\protect\citeauthoryear{{Tingay} et~al.,}{{Tingay}
  et~al.}{2013}]{2013PASA...30....7T}
{Tingay} S.~J.,  et~al., 2013, \mndoi [\pasa] {10.1017/pasa.2012.007}, \href
  {http://adsabs.harvard.edu/abs/2013PASA...30....7T} {30, 7}

\bibitem[\protect\citeauthoryear{{Vernstrom}, {Heald}, {Vazza}, {Galvin},
  {West}, {Locatelli}, {Fornengo}  \& {Pinetti}}{{Vernstrom}
  et~al.}{2021}]{2021MNRAS.505.4178V}
{Vernstrom} T.,  {Heald} G.,  {Vazza} F.,  {Galvin} T.~J.,  {West} J.~L.,
  {Locatelli} N.,  {Fornengo} N.,   {Pinetti} E.,  2021, \mndoi [\mnras]
  {10.1093/mnras/stab1301}, \href
  {https://ui.adsabs.harvard.edu/abs/2021MNRAS.505.4178V} {505, 4178}

\bibitem[\protect\citeauthoryear{Virtanen et~al.,}{Virtanen
  et~al.}{2020}]{2020SciPy-NMeth}
Virtanen P.,  et~al., 2020, \mndoi [Nature Methods]
  {10.1038/s41592-019-0686-2}, \href {https://rdcu.be/b08Wh} {17, 261}

\bibitem[\protect\citeauthoryear{{Wayth} et~al.,}{{Wayth}
  et~al.}{2015}]{2015PASA...32...25W}
{Wayth} R.~B.,  et~al., 2015, \mndoi [\pasa] {10.1017/pasa.2015.26}, \href
  {http://adsabs.harvard.edu/abs/2015PASA...32...25W} {32, e025}

\bibitem[\protect\citeauthoryear{{Wayth} et~al.,}{{Wayth}
  et~al.}{2018}]{2018PASA...35...33W}
{Wayth} R.~B.,  et~al., 2018, \mndoi [\pasa] {10.1017/pasa.2018.37}, \href
  {https://ui.adsabs.harvard.edu/abs/2018PASA...35...33W} {35, e033}

\bibitem[\protect\citeauthoryear{{White} et~al.,}{{White}
  et~al.}{2020a}]{2020PASA...37...17W}
{White} S.~V.,  et~al., 2020a, \mndoi [\pasa] {10.1017/pasa.2020.10}, \href
  {https://ui.adsabs.harvard.edu/abs/2020PASA...37...17W} {37, e017}

\bibitem[\protect\citeauthoryear{{White} et~al.,}{{White}
  et~al.}{2020b}]{2020PASA...37...18W}
{White} S.~V.,  et~al., 2020b, \mndoi [\pasa] {10.1017/pasa.2020.9}, \href
  {https://ui.adsabs.harvard.edu/abs/2020PASA...37...18W} {37, e018}

\bibitem[\protect\citeauthoryear{{de Gasperin} et~al.,}{{de Gasperin}
  et~al.}{2021}]{2021A&A...648A.104D}
{de Gasperin} F.,  et~al., 2021, \mndoi [\aap] {10.1051/0004-6361/202140316},
  \href {https://ui.adsabs.harvard.edu/abs/2021A&A...648A.104D} {648, A104}

\bibitem[\protect\citeauthoryear{{de Gasperin} et~al.,}{{de Gasperin}
  et~al.}{2023}]{2023A&A...673A.165D}
{de Gasperin} F.,  et~al., 2023, \mndoi [\aap] {10.1051/0004-6361/202245389},
  \href {https://ui.adsabs.harvard.edu/abs/2023A&A...673A.165D} {673, A165}

\bibitem[\protect\citeauthoryear{{van Haarlem} et~al.,}{{van Haarlem}
  et~al.}{2013}]{2013A&A...556A...2V}
{van Haarlem} M.~P.,  et~al., 2013, \mndoi [\aap]
  {10.1051/0004-6361/201220873}, \href
  {http://adsabs.harvard.edu/abs/2013A%26A...556A...2V} {556, A2}

\bibitem[\protect\citeauthoryear{{van der Tol}, {Jeffs}  \& {van der
  Veen}}{{van der Tol} et~al.}{2007}]{2007ITSP...55.4497V}
{van der Tol} S.,  {Jeffs} B.~D.,   {van der Veen} A.~J.,  2007, \mndoi [IEEE
  Transactions on Signal Processing] {10.1109/TSP.2007.896243}, \href
  {https://ui.adsabs.harvard.edu/abs/2007ITSP...55.4497V} {55, 4497}

\makeatother
\end{thebibliography}

\end{document}